\documentclass[10pt]{article}
\usepackage{amsmath,amssymb,tikz}
\usepackage{graphicx}
 \allowdisplaybreaks
 \newbox\pippobox
\def\tri {\bigtriangledown}
\def\({\left(} \def\){\right)}
\def\[{\left[} \def\]{\right]}

\newcommand{\be}{\begin{equation}}
\newcommand{\ee}{\end{equation}}
\newcommand{\bea}{\begin{eqnarray}}
\newcommand{\eea}{\end{eqnarray}}
\newcommand{\ba}{\begin{eqnarray}}
\newcommand{\ea}{\end{eqnarray}}

\newcommand{\beq}{\begin{equation}}
\newcommand{\eeq}{\end{equation}}
\newcommand{\beqa}{\begin{eqnarray}}
\newcommand{\eeqa}{\end{eqnarray}}
\newcommand{\beqar}{\begin{eqnarray*}}
\newcommand{\eeqar}{\end{eqnarray*}}

\newcommand{\mt}[1]{\textrm{\tiny #1}}

\newcommand{\veps}{\varepsilon}

\def\de{\delta}

\long\def\symbolfootnote[#1]#2{\begingroup%
\def\thefootnote{\fnsymbol{footnote}}\footnote[#1]{#2}\endgroup}

\newcommand{\aei}{\it Max Planck Institute for Gravitational Physics
(Albert Einstein Institute)\\ Am M\"uhlenberg 1, 14476 Golm,
Germany}

\newcommand{\itp}{\it Kavli Institute for Theoretical Physics,
Key Laboratory of Frontiers in Theoretical Physics,\\
Institute of Theoretical Physics, Chinese Academy of Sciences, Beijing 100190}

\newcommand{\yitp}{\it Yukawa Institute for Theoretical Physics (YITP),\\
Kyoto University, Kyoto 606-8502, Japan}
\begin{document}
\thispagestyle{empty}
\begin{center}

~\vspace{20pt}

{\Large\bf Holographic Entanglement Entropy for the Most General Higher Derivative Gravity}

\vspace{25pt}

Rong-Xin Miao\symbolfootnote[1]{Email:~\sf rong-xin.miao@aei.mpg.de}, Wu-zhong Guo\symbolfootnote[2]{Email:~\sf wuzhong@itp.ac.cn}${}^\dagger{}$

\vspace{10pt}${}^\ast{}$\aei

\vspace{10pt}${}^\dagger{}$\itp

\vspace{10pt}${}^\dagger{}$\yitp

\vspace{2cm}

\begin{abstract}
The holographic entanglement entropy for the most general higher derivative gravity is investigated. We find a new type of Wald entropy, which appears on entangling surface without the rotational symmetry and reduces to usual Wald entropy on Killing horizon. Furthermore, we obtain a formal formula of HEE for the most general higher derivative gravity and work it out exactly for some squashed cones. As an important application, we derive HEE for gravitational action with one derivative of the curvature when the extrinsic curvature vanishes. We also study some toy models with non-zero extrinsic curvature. We prove that our formula yields the correct universal term of entanglement entropy for 4d CFTs. Furthermore, we solve the puzzle raised by Hung, Myers and Smolkin that the logarithmic term of entanglement entropy derived from Weyl anomaly of CFTs does not match the holographic result even if the extrinsic curvature vanishes. We find that such mismatch comes from the `anomaly of entropy' of the derivative of curvature. After considering such contributions carefully, we resolve the puzzle successfully. In general, we need to fix the splitting problem for the conical metrics in order to derive the holographic entanglement entropy. We find that, at least for Einstein gravity, the splitting problem can be fixed by using equations of motion. How to derive the splittings for higher derivative gravity is a non-trivial and open question. For simplicity, we ignore the splitting problem in this paper and find that it does not affect our main results. 
\end{abstract}

\end{center}

 \newpage

\tableofcontents

\section{Introduction}

In \cite{Ryu1,Ryu2}, Ryu and Takayanagi develop a holographic approach to calculate entanglement entropy (EE) of quantum (conformal) field theories in the context of AdS/CFT correspondence \cite{Maldacena}. For a subsystem $A$ on the boundary, they propose an elegant formula of EE
\begin{eqnarray}\label{TakayanagiHEE}
S_A=\frac{\text{Area of}\  \gamma_A}{4G},
\end{eqnarray}
where $\gamma_A$ is the minimal surface in the bulk whose boundary is given by $\partial A$ and $G$ is the bulk Newton constant. Their formula yields the correct EE for two-dimensional CFTs and satisfies the strong subadditivity of EE \cite{Headrick}
\begin{eqnarray}\label{strongsubadditivity}
S_A+S_B \ge S_{A\cup B}+S_{A\cap B}.
\end{eqnarray}

Recently, the conjecture eq.(\ref{TakayanagiHEE}) was proved by Lewkowycz and Maldacena \cite{Maldacena1}. See also \cite{Fursaev, Casini}  for the proof of Ryu-Takayanagi conjecture.  Besides the gravity side there are also many interesting progress in the field theory side, please refer to \cite{Nozaki:2014hna,He:2014mwa,Bousso:2014sda,Bousso:2014uxa,Rosenhaus:2014woa,Rosenhaus:2014nha,Smolkin} for more details.

The formula of Ryu and Takayanagi applies to quantum field theories dual to Einstein Gravity. Thus the corresponding CFTs have only one independent central charge.  To cover more general field theories, one need to generalize their work to higher derivative gravity. A natural candidate of holographic entanglement entropy (HEE) for higher derivative gravity would be Wald entropy:
\begin{eqnarray}\label{Waldentropy}
S_{\text{Wald}}=-2\pi\int d^dy \sqrt{g}\frac{\delta L}{\delta R_{\mu\nu\rho\sigma}}\epsilon_{\mu\nu}\epsilon_{\rho\sigma}.
\end{eqnarray}
However, as pointed out by Hung, Myers and Smolkin\cite{Hung}, Wald entropy does not give the correct universal logarithmic term of EE for CFTs when the extrinsic curvature is non-zero. For Lovelock gravity, we have another entropy formula: the Jacobson-Myers entropy \cite{Jacobson} which differs from Wald entropy by some extrinsic-curvature terms. It turns out that the Jacobson-Myers entropy \cite{Jacobson} yields the correct CFT results \cite{Hung, Boer}. However, there is no similar entropy formula for general higher derivative gravity. One do not know how to derive HEE from the first principle when the extrinsic curvature appears.

The first breakthrough was made by Fursaev, Patrushev and Solodukhin (FPS)\cite{Solodukhin1}. They develop a regularization procedure to deal with the squashed conical singularities. Using this regularization procedure, they successfully obtain HEE for the curvature-squared gravity. Soon after \cite{Solodukhin1}, another important breakthrough was made by Dong \cite{Dong}. Dong find that, similar to holographic Weyl anomaly, the would-be logarithmic terms also contribute to HEE. Dong call such contribution as the `anomaly of entropy'. For the so-called `general higher derivative gravity' whose action including no derivatives of the curvature $S(g, R)$, Dong derive an elegant formula of HEE:
\begin{eqnarray}\label{Dongentropy}
S_{EE}&=&2\pi\int d^dy \sqrt{g}\big{[}\frac{\partial L}{\partial R_{z\bar{z}z\bar{z}}}+\sum_{\alpha}\big{(}\frac{\partial^2L}{\partial R_{zizl}\partial R_{\bar{z}k\bar{z}l}}\big{)}_{\alpha}\frac{8K_{zij}K_{\bar{z}kl}}{q_{\alpha}+1}\big{]},
\end{eqnarray}
where the first term is Wald entropy and the second term is the anomaly of entropy. Please refer to \cite{Dong} for the definition of $q_{\alpha}$. It should be mentioned that Camps \cite{Camps} also made important contributions in this direction. For recent developments of HEE, please refer to \cite{Bhattacharyya1,Bhattacharyya2,Bhattacharyya3,Balasubramanian,Balasubramanian1,1Myers,Headrick,Cremonini, 1Fursaev1,Erdmenger,Safdi,KallolSen}.

So far, HEE for gravitational actions which include derivatives of the curvature is not known. In this paper, we fill this gap by generalizing Dong's work to `the most general higher derivative gravity' $S(g, R,\nabla R,...)$. We find all the possible would-be logarithmic terms and derive a formal formula of HEE for `the most general higher derivative gravity'.
To get more exact formulas, we focus on gravity theories whose action $S(g, R, \nabla R)$ includes only one derivative of the curvature. A natural guess of HEE for $S(g, R, \nabla R)$ would be Dong's formula eq.(\ref{Dongentropy}) with all $\partial$ be replaced by $\delta$. This is however not the case. Instead, we find that new terms should be added to both Wald entropy and anomaly of entropy even if we replace all $\partial$ by $\delta$.  The generalized Wald entropy for $S(g, R, \nabla R)$ is
\begin{eqnarray}\label{GWaldentropy}
S_{\text{G-Wald}}&=&2\pi\int d^dy \sqrt{g}\big{[}\ \frac{\delta L}{\delta R_{z\bar{z}z\bar{z}}}
+2(\frac{\partial L}{\partial \nabla_zR_{\bar{z}i\bar{z}j}}  K_{\bar{z}ij}+c.c ) \ \big{]}\nonumber\\
&=&2\pi\int d^dy \sqrt{g}\big{[}\ -\frac{\delta L}{\delta R_{\mu\nu\rho\sigma}}\epsilon_{\mu\nu}\epsilon_{\rho\sigma}
+2\frac{\partial L}{\partial \nabla_{\alpha}R_{\mu\rho\nu\sigma}}  K_{\beta\rho\sigma} (n^{\beta}_{\ \mu}n_{\alpha\nu}-\epsilon^{\beta}_{\ \mu}\epsilon_{\alpha\nu})\ \big{]}
 \end{eqnarray}
By `generalized Wald entropy', we means the total entropy minus the anomaly of entropy.
Interestingly, a new term proportional to the extrinsic curvature appears in the generalized Wald entropy. This new term only appears on entangling surface without the rotational symmetry, thus it is consistent with Wald's results on Killing horizon. While for the anomaly of entropy, since the general case is very complicated, we set $K_{aij}=0$ for simplicity. If the anomaly of entropy is just Dong's formula with $\partial$ be replaced by $\delta$, it should vanish after we set $K_{aij}=0$. However, we get
\begin{eqnarray}\label{Anomalyentropy}
S_{\text{Anomaly}}&=&2\pi\int d^dy \sqrt{g}\big{[}64\big{(}\frac{\partial^2L}{\partial \nabla_zR_{zizl}\partial \nabla_{\bar{z}}R_{\bar{z}k\bar{z}l}}\big{)}_{\alpha_1}\frac{Q_{zzij}Q_{\bar{z}\bar{z}kl}}{\beta_{\alpha_1}}\nonumber\\
&+&96 i\big{(}\frac{\partial^2L}{\partial \nabla_zR_{zizl}\partial \nabla_{\bar{z}}R_{\bar{z}z\bar{z}k}}\big{)}_{\alpha_1}\frac{Q_{zzij}V_{\bar{z}k}}{\beta_{\alpha_1}}+c.c\nonumber\\
&+&144\big{(}\frac{\partial^2L}{\partial \nabla_zR_{z\bar{z}zl}\partial \nabla_{\bar{z}}R_{\bar{z}z\bar{z}k}}\big{)}_{\alpha_1}\frac{V_{zl}V_{\bar{z}k}}{\beta_{\alpha_1}}
\big{]}.
\end{eqnarray}
Applying the above formula, we resolve the puzzle raised by Huang, Myers and Smolkin (HMS) that the logarithmic term of EE derived from Weyl anomaly of CFTs does not match the holographic result even if the extrinsic curvature vanishes \cite{Hung}. We find that such mismatch comes from the contributions of the derivative of the curvature. After considering these contributions carefully by using the above formula, we resolve the HMS puzzle successfully.

For non-zero extrinsic curvature, we investigate a toy model with Lagrangian $L=\lambda_1\nabla_{\alpha}R\nabla^{\alpha}R+\lambda_2\nabla_{\alpha}R_{\mu\nu}\nabla^{\alpha}R^{\mu\nu}+\lambda_3\nabla_{\alpha}R_{\mu\nu\rho\sigma}\nabla^{\alpha}R^{\mu\nu\rho\sigma}$. We derive HEE and prove it yields the correct logarithmic terms of EE for 4d CFTs.

The paper is organized as follows. In Sect. 2, we briefly review Dong's derivation of HEE for `general higher derivative gravity'. In Sect. 3, we generalize Dong's method to the most general cases.  We obtain a formal formula of HEE for the most general higher derivative gravity. As an exercise, we work out the exact formula for some interesting conical metrics. In Sect. 4, we prove that our formula yields the correct logarithmic term of EE for 4d CFTs. In Sect. 5, we resolve the HMS puzzle. We derive the logarithmic term of entanglement entropy for 6d CFTs from Weyl anomaly and find it is consistent with the holographic result for entangling surfaces with zero extrinsic curvature but without rotational symmetry. In Sect. 6, we compare with our resolution of the HMS puzzle with the one of \cite{Astaneh1,Astaneh2}. Finally, we conclude in Sect 7.

Note added: After this work is finished, there appears two related papers \cite{Astaneh1,Astaneh2}. The authors of \cite{Astaneh1,Astaneh2} claim that convariant total derivatives may contribute to non-trivial entropy and propose to use the entropy of total derivatives to explain the HMS mismatch \cite{Hung}. We notice that their results are based on the FPS regularizations \cite{Solodukhin1}. By applying the Lewkowycz-Maldacena (LM) regularization \cite{Maldacena1, Dong} instead, it is found that the entropy of convariant total derivatives is indeed trivial \cite{Dong2}. In this paper, we use the LM regularization \cite{Maldacena1, Dong} to investigate the HEE.

\section{Dong's proposal of HEE for higher derivative gravity}

In this section, we briefly review Dong's derivation of HEE for higher derivative gravity \cite{Dong}. The key observation of Dong is that, similar to the holographic Weyl anomaly, the would-be logarithmic term also contributes to HEE. As a result, corrections of entropy from the extrinsic curvature emerge:
\begin{eqnarray}\label{coentropy}
\delta S&=&32\pi\int d^dy \sqrt{g} \big{(}\frac{\partial^2L}{\partial R_{zizl}\partial R_{\bar{z}k\bar{z}l}}\big{)}_{\alpha_1}\frac{K_{zij}K_{\bar{z}kl}}{\beta_{\alpha_1}}.
\end{eqnarray}
Dong calls such corrections as the anomaly of entropy. For simplicity, he focuses on the gravity theories without derivatives of the curvature, $S=S(g, R)$. We review Dong's derivation of HEE in this section and generalize it to the most general case $S=S(g, R, \nabla R, ...)$ in the next section.

\subsection{The replica trick}

A useful method to derive HEE is by applying the replica trick. Let us take Einstein Gravity as an example. Recall that the Renyi entropy is defined as
\begin{eqnarray}\label{renyi}
&&S_n=- \frac{1}{n-1}\log tr[\rho^n]=-\frac{1}{n-1}(\log Z_n-n \log Z_1)\\
&&Z_n=Tr[\hat{\rho}^n],\ \rho=\frac{\hat{\rho}}{Tr[\hat{\rho}]},
\end{eqnarray}
where $Z_n$ is the partition function of the field theory on a suitable manifold $M_n$ known as the $\text{n-fold}$ cover.

For theories with a holographic dual we can build a suitable bulk solution $B_n$ whose boundary is $M_n$. Then the gauge-gravity duality identifies the field theory partition function on $M_n$ with the on-shell bulk action on $B_n$
\begin{eqnarray}\label{Zn}
Z_n=Z[M_n]=e^{-S[B_n]}.
\end{eqnarray}
We can derive the HEE by taking the limit $n\to 1$ of Renyi entropy
\begin{eqnarray}\label{SHEE}
S_{EE}&=&\lim_{n\to 1}S_{n}=-\partial_n(\log Tr[\rho^n])|_{n\to 1}=-Tr[\rho\log \rho]\nonumber\\
&=&-\partial_n(\log Z_n-n \log Z_1)|_{n\to 1}=\partial_n(S[B_n]-nS[B_1])|_{n\to 1}\nonumber\\
&=&-\partial_{\epsilon} S_{reg},
\end{eqnarray}
where $S_{reg}=(nS[B_1]-S[B_n])$ is the regularized action and $\epsilon=1-\frac{1}{n}$.
For Einstein gravity, we have
\begin{eqnarray}\label{SEinstein}
S_{reg}=\frac{1}{16\pi G}\int_{Reg \ Cone} dx^{D} \sqrt{G}\  R= \epsilon \  \frac{\text{Area}}{4G}.
\end{eqnarray}
Then we can derive HEE of Einstein gravity as $S=-\frac{\text{Area}}{4G}$.
Note that we work in the Euclidean signature. So entropy formula differs from its usual Lorentzian form by a minus sign.

There is still one question need to be answered. On which surface shall we apply this formula? We know the answer is the minimal surface for Einstein gravity. In general,  according to \cite{Maldacena1}, we require that the analytically continued solution satisfies the linearized equations of motion near the cone $\rho=0$. We call this method the ``boundary condition method''.
The metric of regularized cone is
\begin{eqnarray}\label{metric1}
ds^2=e^{2A}dzd\bar{z}+(g_{ij}+2z K_{zij}+2\bar{z}K_{\bar{z}ij})dy^idy^j +o(\rho^2),
\end{eqnarray}
where  $z=\rho e^{i \tau}$, $dzd\bar{z}=d\rho^2+\rho^2d\tau^2$, $A=-\epsilon\log(\rho)$ and $K_z$ is the extrinsic curvature.
Let us compute the linearized equations of motion $\delta{G_{zz}}=8\pi G \delta T_{zz}$. We focus on the divergent terms, going like $1/\rho$ near the origin. Since the stress tensor is not expected to be singular, we have
\begin{eqnarray}
\delta R_{zz}=-\frac{\epsilon}{z} K_z+ \text{regular terms}.
\end{eqnarray}
Requiring the above equation to be regular near the cone, we get $K_z=K_{\bar{z}}=0$. This is just the condition of the minimal surface.

There is another method to derive the minimal surface conditions. We call it the `cosmic brane method'.  Consider the action
\begin{eqnarray}
S_{total}=S_{EH}+S_B=-\frac{1}{16\pi G_N}\int_{Reg} d^{D}x\sqrt{G}R+\frac{\epsilon}{4G_N}\int d^{D-2}y\sqrt{g}.
\end{eqnarray}
In the limit $\epsilon\to 0$, we can treat $S_B$ as the action of a probe brane and find its location by minimizing $S_B$ without back reaction on the bulk fields. This gives exactly the minimal surface.

We have shown how to derive HEE for Einstein Gravity and how to derive the location of the cone. Now let us try to generalize it to higher derivative gravity.

\subsection{Would-be logarithmic terms}
According to \cite{Dong}, the metric of regularized cone is
\begin{eqnarray}\label{metric}
ds^2=e^{2A}[dzd\bar{z}+e^{2A}T(\bar{z}dz-zd\bar{z})^2]+\big{(} g_{ij}+2K_{aij}x^a+Q_{abij}x^ax^b\big{)}dy^idy^j\nonumber\\
+2i e^{2A}(U_i+V_{ai}x^a)(\bar{z}dz-zd\bar{z})dy^i+...,
\end{eqnarray}
where  $T,g_{ij}, K_{aij},Q_{abij},U_i,V_{ai}$ are independent of z and $\bar{z}$, with the exception that $Q_{z\bar{z}ij}=Q_{\bar{z}zij}$ contains a factor $e^{2A}$. The warp factor $A$ is regularized by a thickness parameter $a$ as $A=-\frac{\epsilon}{2}\lg(z\bar{z}+a^2)$. As we shall show below, the result is independent of the choice of regularization.

The key observation of \cite{Dong} is that
\begin{eqnarray}\label{Key}
\int \rho d\rho \partial_z A\partial_{\bar{z}} A e^{-\beta A}=-\frac{\epsilon}{4\beta},
\end{eqnarray}
where $z=\rho e^{i\tau}$. Naively the left hand of eq.(\ref{Key}) is in order $o(\epsilon^2)$. Magically it becomes in order $o(\epsilon)$ after regularization. The magic happens because  would-be logarithmic divergence gets a $\frac{1}{\epsilon}$ enhancement:
\begin{eqnarray}\label{Key1}
\int d\rho \frac{1}{\rho^{1-\beta \epsilon}}  \sim \frac{1}{\beta \epsilon}.
\end{eqnarray}
As we know, the coefficient of a would-be logarithmic divergence is universal (like anomaly). So eq.(\ref{Key}) is independent of the regularization. In fact, we can give a very simple proof. It is known that the following formula is universal
\begin{eqnarray}\label{Key10}
\int dzd\bar{z} e^{-\beta A}\partial_z\partial_{\bar{z}} A =-\pi\epsilon.
\end{eqnarray}
This formula is usually used to derive Wald entropy.
Performing integration by parts, we get
\begin{eqnarray}\label{Key2}
\int dzd\bar{z} e^{-\beta A}\partial_z A \partial_{\bar{z}} A =\frac{-\pi\epsilon}{\beta}.
\end{eqnarray}
which is exactly eq.(\ref{Key}). It should be mentioned that we can drop the boundary terms safely. One can check that the boundary term is zero after regularization. Note that eqs.(\ref{Key},\ref{Key10}) are only true to linear order in$\epsilon$. We ignore the higher-order terms because they do not contribute to the HEE.

\subsection{Dong's formula: HEE for four-derivative gravity}

Now let us focus on the four-derivative gravity whose action $S(g, R)$ contains no derivatives of the curvature. By four-derivative gravity, we means the equations of motion are four order differential equations. This is the case investigated in \cite{Dong}. From the regularized metric eq.(\ref{metric}), we can derive the curvature with non-vanishing derivatives of $A$ as
\begin{eqnarray}\label{curvature1}
&&R_{z\bar{z}z\bar{z}}=e^{2A}\partial_z\partial_{\bar{z}}A+...,\nonumber\\
&&R_{zizj}=2K_{zij}\partial_{z}A+...,\nonumber\\
&&R_{z\bar{z}zi}=i e^{2A} U_i \partial_{z}(z\partial_{z}A)+...,
\end{eqnarray}
where ``...'' denotes terms without derivatives of $A$. One can get the other curvatures by  exchanging $z,\bar{z},i,j$ and complex conjugate. For the reason will be clear in sect. 3, $R_{z\bar{z}zi}\sim e^{2A} U_i \partial_{z}(z\partial_{z}A)\sim 0$ actually does not contribute to HEE. Thus, from eqs.(\ref{curvature1},\ref{Key10},\ref{Key2}), we can derive the HEE as
\begin{eqnarray}\label{HEE1}
S_{EE}&=&2\pi\int d^dy \sqrt{g}\big{[}\frac{\partial L}{\partial R_{z\bar{z}z\bar{z}}}+16\big{(}\frac{\partial^2L}{\partial R_{zizl}\partial R_{\bar{z}k\bar{z}l}}\big{)}_{\alpha_1}\frac{K_{zij}K_{\bar{z}kl}}{\beta_{\alpha_1}}
\big{]}
\end{eqnarray}
The first term above is just the Wald entropy, and the second term denotes the anomly of entropy \cite{Dong}. It should be stressed that, unlike $K_{aij}$, $U_i$ could not appear in the formula of HEE eq.(\ref{HEE1}). Otherwise, it would yield wrong results of entropy for stationary black holes. As we shall show in next section, $R_{z\bar{z}zi}=i e^{2A} U_i \partial_{z}(z\partial_{z}A)$ indeed do not contribute to HEE.

\section{HEE for the most general higher derivative gravity}

In this section, we investigate HEE for the most general higher derivative gravity. Firstly, we discuss the splitting problems for the conical metrics. Then we find all the possible would-be logarithmic terms and derive a formal formula of HEE for the most general higher derivative gravity. Finally, we work out the formal formula exactly for some special conical metrics.

\subsection{Splitting problems}

The splitting problems appear because we can not distinguish $r^2$ and $r^{2n}$ in the expansions of the conical metrics. That is because $r^2$ and $r^{2n}$ are of the same order in the limit $n\to 1$ when we calculate HEE. It should be mentioned that the splitting problem is ignored in the initial works of Dong and Camps \cite{Dong,Camps}. However they both change their mind and realize the splitting is necessary later \footnote{We thank Dong and Camps for discussions on this problem.}. Recently Camps etal generalize the conical metrics to the case without $Z_n$ symmetry, where the splitting problem appears naturally \cite{Camps1}. Inspired by the works of \cite{Maldacena1,Dong,Camps1}, a natural way to fix the splitting problem is by using equations of motion. As we shall prove below, this is indeed the case at least for Einstein gravity. For the higher derivative gravity, how to fix the splitting problem is a non-trivial and open problem. We leave it for future work. It should be mentioned that the splitting problem does not affect the main results of this paper. We shall explain the reasons briefly at the end of this subsection.

Let us start with the general squashed conical metric \cite{Dong,Camps}
\begin{eqnarray}\label{cone}
ds^2=e^{2A}[dzd\bar{z}+T(\bar{z}dz-zd\bar{z})^2]+2i V_i(\bar{z}dz-zd\bar{z})dy^i\nonumber\\
+( g_{ij}+Q_{ij})dy^idy^j,
\end{eqnarray}
where $g_{ij}$ is the metric on the transverse space and is independent of $z, \bar{z}$. $A=-\frac{\epsilon}{2}\lg(z\bar{z}+a^2)$ is regularized warp factor. $T, V_i, Q_{ij}$ are defined as
\begin{eqnarray}\label{TVQ}
&&T=\sum_{n=0}^{\infty}\sum_{m=0}^{P_{a_1...a_n}+1} e^{2 m A }T_{m\ a_1...a_n}x^{a_1}...x^{a_n},\nonumber\\
&&V_i=\sum_{n=0}^{\infty} \sum_{m=0}^{P_{a_1...a_n}+1} e^{2 m A }V_{m\ a_1...a_n i}x^{a_1}...x^{a_n},\nonumber\\
&&Q_{ij}=\sum_{n=1}^{\infty} \sum_{m=0}^{P_{a_1...a_n}} e^{2 m A }Q_{m\ a_1...a_n ij}x^{a_1}...x^{a_n}.
\end{eqnarray}
Here $z,\bar{z}$ are denoted by $x^{a}$ and $P_{a_1...a_n}$ is the number of pairs of
$z,\bar{z}$ appearing in $a_1...a_n$. For example, we have $P_{zz\bar{z}}=P_{z\bar{z}z}=P_{\bar{z}zz}=1$, $P_{z\bar{z}z\bar{z}}=2$ and $P_{zz...z}=0$.
Expanding $T,V,Q$ to the first few terms in Dong's notations, we have
\begin{eqnarray}\label{TVQ1}
&&T=T_0+e^{2A}T_{1}+O(x),\nonumber\\
&&V_i=U_{0 \ i}+e^{2A}U_{1\ i}+O(x),\nonumber\\
&&Q_{ij}=2K_{aij}x^a+Q_{0\ abij}x^ax^b+2e^{2A}Q_{1\ z\bar{z}ij}\ z \bar{z}+O(x^3)
\end{eqnarray}
How to split $W$ ($W$ denote $T, V, Q$) into $\{W_{0}, W_{1}, ..., W_{P+1}\}$ is an important problem. Inspired by \cite{Maldacena1},  it is expected that the splitting problem can be fixed by equations of motion. Let us take Einstein gravity in vacumm as an example.  We denote the quations of motion by $E_{\mu\nu}=R_{\mu\nu}-\frac{R-2\Lambda}{2}G_{\mu\nu}=0$ . Focus on terms which are important near $x^a=0$, we have
\begin{eqnarray}\label{curvature}
R_{ab}&=&2K_{(a}\nabla_{b)}A -g_{ab}K^c\nabla_cA+e^{2A}[(12T_1+4 U^2)g_{ab}-Q_{1\ abi}^{\ \ \ \ \ \ i}]\nonumber\\
&&+K_{aij}K_{b}^{\ ij}+(12T_0+8 U_{0}U_{1}) g_{ab}-Q_{0\ abi}^{\ \ \ \ \ \  i}\nonumber\\
R_{ai}&=&3\varepsilon_{ba}V^{b}_{\ i}+D^mK_{ami}-D_i K_a, \nonumber\\
R_{ij}&=&r_{ij}+8U_iU_j-Q_{1\ aij}^{\ a}+e^{-2A}[2K_{aim}K^{am}_{\ \ \ j}-K^aK_{aij}+16 U_{0\ (i}U_{1\ j)}-Q_{0\ aij}^{\ a}],\nonumber\\
R&=&r+16U^2+24T_1-2Q_{1\ a\ i}^{\ a\ i}+e^{-2A}(3K_{aij}K^{aij}-K^aK_a+24T_0-2Q_{0\ a\ i}^{\ a\ i}+32U_0U_1),
\end{eqnarray}
where $A=-\frac{\epsilon}{2} \log z\bar{z} $, $\varepsilon_{z\bar{z}}=\frac{i}{2}$ and $g_{z\bar{z}}=\frac{1}{2}$. Let us firstly consider the leading term of $E_{zz}$, we get
\begin{eqnarray}\label{Ezz}
E_{zz}=2K_z \nabla_z+...=-\epsilon \frac{K_z}{z}+...=0.
\end{eqnarray}
Requiring the above equation to be regular near the cone, we obtain the minimal surface condition $K_z=K_{\bar{z}}=0$ \cite{Maldacena1}. To derive $T_0$ and $Q_0$, we need consider the subleading terms of $E_{z\bar{z}}, E_{ij}$ and $E_{\mu}^{\mu}$. We have
\begin{eqnarray}\label{Eij}
&&E_{z\bar{z}}=e^{2A}(...)+[Q_{0\ z\bar{z}i}^{\ \ \ i}-2K_{zij}K_{\bar{z}}^{\ ij}+K_z K_{\bar{z}}-4U_0 U_1]=0,\nonumber\\
&&E_{ij}=(...)+e^{-2A} [2K_{aim}K^{am}_{\ \ \ j}-K^aK_{aij}+16 U_{0\ (i}U_{1\ j)} -Q_{0\ aij}^{\ a}\nonumber\\
&&\ \ \ \ \ \ \ \ -\frac{1}{2}g_{ij}
(3K_{aij}K^{aij}-K^aK_a+24T_0-2Q_{0\ a\ i}^{\ a\ i}+32U_0U_1) ]=0,\nonumber\\
&&E_{\ \mu}^{\mu}=(...)+\frac{2-D}{2}e^{-2A}[3K_{aij}K^{aij}-K^aK_a+24T_0-2Q_{0\ a\ i}^{\ a\ i}+32U_0U_1]=0.
\end{eqnarray}
Here $ (...)$ denote the leading terms which can be used to determine $T_1, U_{1 i}, Q_{1 z\bar{z}ij}$ and $g_{ij}$. From the subleading terms of the above equations, we can derive a unique solution
\begin{eqnarray}\label{splittingEinstein}
&&T_0=\frac{1}{24}(K_{aij}K^{aij}-K_aK^a),\nonumber\\
&&Q_{0z\bar{z}ij}=(K_{zim}K_{zj}^{\ \ m}-\frac{1}{2}K_z K_{\bar{z}ij}+c.c.)+4 U_{0\ (i}U_{1\ j)}
\end{eqnarray}
Now we have fixed the splitting of $T$ and $Q_{z\bar{z}ij}$ by using equations of motion. Note that Einstein equations does not fix $U_{0\ i}$. That is not surprising. $U_i$ can be regarded as the `gauge fields' which are related to the coordinate transformations \cite{Camps}. It is clear that equations of motion can not fix the gauge fields completely. It should be mentioned that, if we reqiure that a special background metric such as AdS is a solution, then different theories of gravity can share the same splittings. That is because we have imposed additional conditions. Recently, the splittings eq.(\ref{splittingEinstein}) are used to derive the universal terms of entanglement entropy for 6d CFTs \cite{Miao2}. It turns out that eq.(\ref{splittingEinstein}) is the necessary condition that all the theories of higher derivative gravity with an AdS solution yield the consistent results for the universal terms of entanglement entropy \cite{Miao2}.

In addition to equations of motion, there is another principle which may help us to get some insights into the splitting problem. The entropy should reduce to Wald entropy in stationary spacetime. We call this principle as the `stationary principle'.  Let us take $\nabla_{\mu}R_{\nu\rho\sigma\alpha}\nabla^{\mu}R^{\nu\rho\sigma\alpha}$ as an example. In stationary spacetime, we have $K_{aij}=Q_{zzij}=Q_{\bar{z}\bar{z}ij}=0$. Applying the method will be developed in the next section, we can derive the HEE as
\begin{eqnarray}\label{ConstraintWald}
S_{HEE}=S_{Wald}+\int dy^{D-2}\sqrt{g} 128\pi ( Q_{0z\bar{z}ij}Q_{0z\bar{z}}^{\  \ ij}+9 T_0^2+5 (U_{0\ i}U_0^{\ i})^2+ \text{mixed terms of }T_0, Q_0, U_0 ).
\end{eqnarray}
To be consistent with Wald entropy, we must have $T_0=U_{0\ i}=Q_{0z\bar{z}ij}=0$ in stationary spacetime. This implies that $T_0, U_{0\ i}$ and $Q_{0z\bar{z}ij}$ should be either zero or functions of the extrinsic curvatures.  This is indeed the case for the splitting eqs.(\ref{splittingEinstein}). The `stationary principle' tells us that the splitting problem disappears if we focus on the cases with zero extrinsic curvature.  By dimensional analysis, we note that $U_{0\ i}\sim O(K)$. However, it is impossible to express $U_{0\ i}$ in terms of the extrinsic curvature $K_{aij}$. Thus, a natural choice would be $U_{0\ i}=0$.

In this paper, for simplicity, we keep only the highest order of $T, V_i, Q_{ij}$ eq.(\ref{TVQ}) to illustrate our approach. This is also the case studied in \cite{Dong}. In other words, we ignore the splittting probelm in most parts of this paper. For example, we set $T_0=U_0=Q_{0}=0$ when we investigate the entropy of higher derivative gravity $S(g, R, \nabla R)$.  According to the `stationary principle', equivalently, we have zero extrinsic curvatures. It should be mentioned that this condition $T_0=U_0=Q_{0}=0$ does not affect our main results (eqs.(\ref{GWaldentropy},\ref{Anomalyentropy},\ref{dA},\ref{dAdA}) and the results in Sect. 4 and Sect. 5). Straightforward calculations can show that $T_0, U_{0\ i}, Q_{0\ z\bar{z}ij}$ do not contribute to the generalized Wald entropy eq.(\ref{GWaldentropy}). However, they indeed appear in the anomaly of entropy, see the appendix. Recall that eq.(\ref{Anomalyentropy}) is derived under the condition $K_{aij}=0$. According to the `stationary principle', the condition $K_{aij}=0$ yields $T_0=U_{0\ i}= Q_{0\ z\bar{z}ij}=0$. Thus $T_0, U_{0\ i}, Q_{0\ z\bar{z}ij}$ does not affect eq.(\ref{Anomalyentropy}). Because we only use eq.(\ref{Anomalyentropy}) to resolve the HMS puzzle (the HMS puzzle is found under the condition $K_{aij}=0$), so $T_0, U_{0\ i}, Q_{0\ z\bar{z}ij}$ does not affect our resolution of the HMS puzzle in Sect. 5. As we shall show in Sect. 4, only the leading terms $T_1=-\frac{1}{12}, Q_{1\ z\bar{z}ij}=\frac{1}{2}G_{ij}$ contribute to the logarithmic term of EE . And the subleading terms $T_0\sim Q_{0\ z\bar{z}ij}\sim o(K^2)$ are irrelevant to the logarithmic term of EE for 4d CFTs in Sect. 4. For the above reasons, the splitting problem does not affect the main results of this paper (eqs.(\ref{GWaldentropy},\ref{Anomalyentropy},\ref{dA},\ref{dAdA}) and the results in Sect. 4 and Sect. 5).

\subsection{General would-be logarithmic terms}

Using the squashed cone metric (\ref{cone}), we can calculate the action of most general higher derivative gravity and then select the relevant terms to derive HEE. Now let us discuss all the possible terms relevant to HEE. The discussions of this subsection are universal and independent of the splitting of the conical metrics.

Let us denote the general derivatives by
\begin{eqnarray}\label{derivatives}
\hat{\partial}=c^{mn}\partial_z^m\partial_{\bar{z}}^n,
\end{eqnarray}
where $c^{mn}$ are arbitrary constants. Since only $o(\epsilon)$ terms contribute to HEE, we only need to consider terms with at most two $A$: $\hat{\partial}A, \hat{\partial}A\hat{\partial}A$. For the first case $\hat{\partial}A$, it is easy to find that only the following terms contribute to HEE

\begin{eqnarray}\label{dA}
\int dz d\bar{z} z^m\bar{z}^n\partial_z^{m+1}\partial_{\bar{z}}^{n+1}A&=&\int dz d\bar{z} (-1)^{m+n}m!n!\partial_z\partial_{\bar{z}}A\nonumber\\
&=&(-1)^{m+n+1} m!n! \pi \epsilon.
\end{eqnarray}
 Equivalently, we have
 \begin{eqnarray}\label{dAD}
\partial_z^{m+1}\partial_{\bar{z}}^{n+1}A=-\pi \epsilon \partial_z^{m}\partial_{\bar{z}}^{n}\bar{\delta}(z,\bar{z}).
\end{eqnarray}
These terms contribute to the Wald entropy. Note that the delta function is defined as $\int dz d\bar{z} \bar{\delta}(z,\bar{z})=1$.

As for the second case $\hat{\partial}A\hat{\partial}A$, we should focus on the would-be logarithmic terms. That is because only such terms could gain a $\frac{1}{\epsilon}$ enhancement. The only possible terms are
\begin{eqnarray}\label{dAdA}
\int dz d\bar{z} z^m\bar{z}^n\partial_z^{m+1}A\partial_{\bar{z}}^{n+1}Ae^{-\beta A}&=&\int dz d\bar{z}(-1)^{m+n}m!n!\partial_zA\partial_{\bar{z}}Ae^{-\beta A}\nonumber\\
&=&(-1)^{m+n+1}m!n!\frac{\pi \epsilon}{\beta} .
\end{eqnarray}
Equivalently, we have
 \begin{eqnarray}\label{dAdAD}
\partial_z^{m+1}A\partial_{\bar{z}}^{n+1}Ae^{-\beta A}=-\frac{\pi \epsilon}{\beta}\partial_z^{m}\partial_{\bar{z}}^{n}\bar{\delta}(z,\bar{z}).
\end{eqnarray}
These terms contribute to the anomaly of entropy. It should be mentioned that eqs.(\ref{dA},\ref{dAdA}) are only true to linear order in$\epsilon$. We ignore the higher-order terms because they do not contribute to the HEE.

The simplest method to prove eq.(\ref{dAdA}) is by applying integration by part and dropping the irrelevant terms such as $\hat{\partial}\partial_z\partial_{\hat{z}}A \hat {\partial}A, \hat{\partial}A \hat {\partial}A\hat{\partial}A$ and so on. This is the method we used in eq.(\ref{dAdA}). We can also prove eq.(\ref{dAdA}) by using Dong's method. Recall that $A=-\frac{\epsilon}{2}\log(z\bar{z})$, we have  $z^m\partial_z^{m+1}A=-\frac{\epsilon}{2}(-1)^m\frac{m!}{z}$. Thus we can derive
\begin{eqnarray}\label{dAdA1}
\int \rho d\rho z^m\bar{z}^n\partial_z^{m+1}A\partial_{\bar{z}}^{n+1}Ae^{-\beta A}&=&\int  d\rho(-1)^{m+n}m!n!\frac{\epsilon^2}{4}\rho^{-1+\beta \epsilon}\nonumber\\
&=&(-1)^{m+n}m!n!\frac{\epsilon}{4\beta}\rho^{\beta \epsilon}|^{\infty}_{0}\nonumber\\
& \cong &(-1)^{m+n+1}\frac{\epsilon}{4\beta} m!n!.
\end{eqnarray}
Here $\cong$ denotes equivalence after regularization. For simplicity, the above equation is illustrated in a regularization-independent way. Now let us use Dong's regularization with $A=-\frac{\epsilon}{2}\log(z\bar{z}+a^2)$ to rederive it. We have
 \begin{eqnarray}\label{dAdA2}
\int \rho d\rho z^m\bar{z}^n\partial_z^{m+1}A\partial_{\bar{z}}^{n+1}Ae^{-\beta A}&=&\int  d\rho(-1)^{m+n}m!n!\frac{\epsilon^2}{4}\frac{\rho^{3+2m+2n}}{(\rho^2+a^2)^{2+m+n-\frac{\beta\epsilon}{2}}}\nonumber\\
&=&(-1)^{m+n} m!n!a^{\beta  \epsilon} \frac{\epsilon^2 \Gamma \left(-\frac{1}{2} \epsilon \beta \right) \Gamma (2+m+n)}{8 \Gamma \left(2+m+n-\frac{\epsilon \beta }{2}\right)}\nonumber\\
&= &(-1)^{m+n+1}\frac{\epsilon}{4\beta} m!n!+O(\epsilon^2).
\end{eqnarray}
Following \cite{Dong}, we set $a$ finite so that $a^{\beta \epsilon}=1+O(\epsilon)$. We have also used $\Gamma[\epsilon]=\frac{1}{\epsilon}+O(\epsilon^0)$ in the above derivations.

It should be stressed that terms contains $\partial_z\partial_{\bar{z}}A$, $z\partial_z A$ or $\bar{z}\partial_{\bar{z}}A $ in the second case would not contribute to HEE,
\begin{eqnarray}\label{ddAdA}
&&\hat{\partial}\partial_z\partial_{\hat{z}}A \hat {\partial}A=0,\nonumber\\
&&\hat{\partial}(z\partial_zA) \hat {\partial}A=\hat{\partial}(\bar{z}\partial_{\bar{z}}A) \hat {\partial}A=0.\nonumber\
\end{eqnarray}
That is because $\partial_z\partial_{\bar{z}}A =-\frac{\epsilon}{2} \frac{a^2}{(a^2+r^2)^2}$, so $\hat{\partial}\partial_z\partial_{\hat{z}}A \hat {\partial}A$ at least in order $\epsilon^2 a^2$. Note that $(a^2+r^2)$ always appear as a whole in the denominator. To cancel $a^2$, we must have $\epsilon^2 \frac{a^2}{a^2+r^2}$ after integration. However this is a $r^{-2}$ term rather than a would be logarithmic term $\frac{1}{\epsilon}r^{-\epsilon}$. So we can not cancel $\epsilon$ and $a^2$ at the same time. Similar for the second case, $\hat{\partial}(z\partial_zA) \hat {\partial}A$ is also at least in order $\epsilon^2 a^2$. Thus, it does not contribute to HEE either.
Maybe the most quick way to see that $\partial_z\partial_{\bar{z}}A$ and $z\partial_z A$ do not contribute to HEE is by identifying $A=-\frac{\epsilon}{2}\log(z\bar{z})$. So we have $\hat{\partial}\partial_z\partial_{\bar{z}}A=\hat{\partial}(z\partial_z A)=0$, which can not contribute to HEE at all.

Using eqs.(\ref{dA},\ref{dAdA}), we can derive HEE for most general higher derivative gravity as
\begin{eqnarray}\label{mostgeneralHEE}
S_{HEE}&=&-\partial_{\epsilon} S_{reg}|_{\epsilon=0}\nonumber\\
&=&2\pi \delta(z,\bar{z})\hat{g}^{ab}\big{(}\frac{\delta S}{\delta \partial_a \partial_b A}+\frac{1}{\beta_{\alpha}}[\frac{\delta}{\delta \partial_b A}(\frac{\delta S}{\delta \partial_a A }|_{\partial_z\partial_{\bar{z}}A=0})]_{\alpha}\big{)}|_{\epsilon=0},
\end{eqnarray}
where a sum over $\alpha$ is implied. Note that, we drop all the $\hat{\partial}\partial_z\partial_{\bar{z}}A$ terms after one variation of  $\partial_a A$ in the second term of eq.(\ref{mostgeneralHEE}). This formula applies to the most general higher derivative gravity. It is one of the main results of this paper.  Let us comment on our formula (\ref{mostgeneralHEE}).

Firstly, the first term of eq.(\ref{mostgeneralHEE}) is the generalized Wald entropy. It should be stressed that not only $R_{z\bar{z}z\bar{z}}$ and its covariant derivative $\nabla^n R_{z\bar{z}z\bar{z}}$ but aslo many other terms may contribute to the generalized Wald entropy.  For example, we have
\begin{eqnarray}\label{Waldlike}
\nabla_zR_{\bar{z}i\bar{z}j}=K_{\bar{z}ij}\partial_z\partial_{\bar{z}}A+....
\end{eqnarray}
Clearly, the above term contributes to the generalized Wald entropy and is not included in the usual Wald entropy $\frac{\delta S}{\delta R_{\mu\nu\rho\sigma}}\epsilon^{\mu\nu}\epsilon^{\rho\sigma}$. Note that such new generalized Wald entropy appears only in the dynamic space-time. Thus nothing goes wrong with Wald's formula which is designed for the stationary black holes. We shall discuss the generalized Wald entropy in details in the next subsection.

Secondly, the second term of eq.(\ref{mostgeneralHEE}) is the anomaly of entropy. In general, it is very difficult to calculate such terms for the most general higher derivative gravity. Let us play a trick.  Setting $A=-\frac{\epsilon}{2}\log[z\bar{z}]$ and keeping only the would-be logarithmic term  $\frac{1}{2} dzd\bar{z}e^{-\beta A}\frac{\epsilon^2}{z\bar{z}}$ in the action, then replacing it by $\frac{2\pi}{\beta}$, we obtain the final result.
\begin{eqnarray}\label{mostgeneralAnomaly}
&&S_{\text{Action}}=\int \frac{1}{2} dzd\bar{z}\sum_{\alpha}C_{\alpha}e^{-\beta_{\alpha}A}\frac{\epsilon^2}{z\bar{z}}+...\nonumber\\
&&S_{\text{Anomaly of entropy}}=\sum_{\alpha}C_{\alpha}\frac{2\pi}{\beta_{\alpha}}.
\end{eqnarray}

Thirdly, we have found all the relevant terms with HEE in order $O(A)$ and $O(A^2)$. A natural question is whether terms in higher order $O(A^{n+2})$ contribute to HEE or not. In general, only would-be $(\log \rho)^{n+1}$ terms may get an enhancement after regularization. Let us discuss these terms briefly. Recall that we have
\begin{eqnarray}\label{Key3}
e^{-\beta A}\partial_z A\partial_{\bar{z}} A=\frac{-\pi\epsilon}{\beta} \delta (z,\bar{z}).
\end{eqnarray}
Taking the derivatives of the above equation by $\beta$, we can derive
\begin{eqnarray}\label{Key4}
A^ne^{-\beta A}\partial_z A\partial_{\bar{z}} A=\frac{-\pi n!\epsilon}{\beta^{n+1}} \delta (z,\bar{z}).
\end{eqnarray}
Naively, the left hand side of eq.(\ref{Key4}) is in order $o(\epsilon^{n+2})$. However it becomes in order  $o(\epsilon)$ after regularization. Actually, this is the would be $(\log \rho)^{n+1}$ terms. This kind of terms may contribute to HEE for some crazy regularized cone metrics. However, if we focus on higher derivative gravity with the regularized cone eq.(\ref{cone}) , only eq.(\ref{dAdA}) is already enough. That is because the factor $e^{\beta A}$ always appear as an entirety in the regularized metric and the action \cite{Dong}, and $A^n \hat{\partial}A\hat{\partial}A$ terms never appear separately.  Thus only the would-be logarithmic term contribute to HEE of higher derivative gravity. Based on eqs.(\ref{dA},\ref{dAdA}), in Sect.4 we shall prove that our formulas of HEE yield the correct universal logarithmic terms of EE for 4d CFTs. This can be regarded as a support of the fact that terms in higher order $O(A^{n+2})$ do not contribute to HEE.

To summary, we have found all the would-be logarithmic terms and obtained a formal formula of HEE for the most general higher derivative gravity.  In the next section, we shall work out this formula exactly for some squashed cone metrics.

\subsection{HEE for six-derivative gravity}

In this subsection, we investigate HEE of six-derivative gravity. By six-derivative gravity, we mean the equations of motion are six order differential equations. Its action can always be rewritten in the form $S(g, R, \nabla R)$. We firstly derive the generalized Wald entropy for the general cone metric and then calculate the anomaly of entropy for some special cone metric.

Let us firstly investigate the generalized Wald entropy. It come from the first term of eq.(\ref{mostgeneralHEE}). As we have mentioned in the above section, in addition to $R_{z\bar{z}z\bar{z}}$ and its covariant derivative $\nabla_{\mu} R_{z\bar{z}z\bar{z}}$, many other terms may contribute to the generalized Wald entropy. We list all the possible terms relevant to the generalized Wald entropy below.
\begin{eqnarray}\label{Waldliketerms}
&&R_{z\bar{z}z\bar{z}}=e^{2A}\partial_z\partial_{\bar{z}}A+...,\nonumber\\
&&\nabla_zR_{z\bar{z}z\bar{z}}=e^{2A}\partial_z^2\partial_{\bar{z}}A+...,\nonumber\\
&&\nabla_zR_{\bar{z}z\bar{z}i}=2i U_i e^{2A}\partial_z\partial_{\bar{z}}A+...,\nonumber\\
&&\nabla_iR_{z\bar{z}zj}=2K_{\bar{z}ij}\partial_z\partial_{\bar{z}}A+...,\nonumber\\
&&\nabla_zR_{\bar{z}i\bar{z}j}=2K_{\bar{z}ij}\partial_z\partial_{\bar{z}}A+....
\end{eqnarray}
Using the above formulae, we can derive
\begin{eqnarray}\label{Waldsix}
S_{\text{G-Wald}}=2\pi\int d^dy \sqrt{g}\big{[}&&\frac{\partial L}{\partial R_{z\bar{z}z\bar{z}}}\nonumber\\
&-&\frac{1}{\sqrt{g}}\partial_z(\sqrt{g}\frac{\partial L}{\partial \nabla_zR_{\bar{z}z\bar{z}z}}) +c.c\nonumber\\
&+&4i\frac{\partial L}{\partial \nabla_zR_{\bar{z}z\bar{z}i}}  U_i+c.c\nonumber\\
&+&2\frac{\partial L}{\partial \nabla_zR_{\bar{z}i\bar{z}j}}  K_{\bar{z}ij}+c.c\nonumber\\
&+&4\frac{\partial L}{\partial \nabla_iR_{z\bar{z}zj}}  K_{zij}+c.c\ \
\big{]}.
\end{eqnarray}
Take into account that $\Gamma^{z}_{zi}=-2i U_i, \Gamma^{i}_{jz}=K_{z j}^{\ i}, \Gamma^{z}_{ij}=-2 K_{\bar{z}ij}$, we obtain the generalized Wald entropy as
\begin{eqnarray}\label{Waldsix1}
S_{\text{G-Wald}}&=&2\pi\int d^dy \sqrt{g}\big{[}\ \frac{\partial L}{\partial R_{z\bar{z}z\bar{z}}}-\nabla_{\mu}\frac{\partial L}{\partial\nabla_{\mu} R_{z\bar{z}z\bar{z}}}
+2(\frac{\partial L}{\partial \nabla_zR_{\bar{z}i\bar{z}j}}  K_{\bar{z}ij}+c.c ) \ \big{]}\nonumber\\
&=&2\pi\int d^dy \sqrt{g}\big{[}\ \frac{\delta L}{\delta R_{z\bar{z}z\bar{z}}}
+2(\frac{\partial L}{\partial \nabla_zR_{\bar{z}i\bar{z}j}}  K_{\bar{z}ij}+c.c ) \ \big{]}.
 \end{eqnarray}
Remarkably, a new term proportional to the extrinsic curvature $K_{aij}$ appears in the generalized Wald entropy. This new term vanishes for stationary black holes and thus is consistent with Wald's results. In general, self conjugate terms such as $T, U_i, Q_{z\bar{z}ij}...$ could not contribute new terms to the generalized Wald entropy, otherwise it conflicts with Wald entropy for stationary black holes. That is because, in general, these self conjugate terms are non-zero in stationary spacetime. Indeed,  $T, U_i,Q_{z\bar{z}ij}$ do not appear in our generalized Wald entropy eq.(\ref{Waldsix1}) for six-derivative gravity. The above generalized Wald entropy can be written in a covariant form as
\begin{eqnarray}\label{Waldsix2}
S_{\text{G-Wald}}=2\pi\int d^dy \sqrt{g}\big{[}\ -\frac{\delta L}{\delta R_{\mu\nu\rho\sigma}}\epsilon_{\mu\nu}\epsilon_{\rho\sigma}
+2\frac{\partial L}{\partial \nabla_{\alpha}R_{\mu\rho\nu\sigma}}  K_{\beta\rho\sigma} (n^{\beta}_{\ \mu}n_{\alpha\nu}-\epsilon^{\beta}_{\ \mu}\epsilon_{\alpha\nu})\ \big{]}.
 \end{eqnarray}
 It should be mentioned that the extrinsic curvature flips the sign under $n_{\mu a}\to -n_{\mu a}$ ($a$ denotes the flat index and $\mu$ is the spacetime index). So it seems that the generalized Wald entropy eqs.(\ref{Waldsix1},\ref{Waldsix2}) depend on the orientation of the surface. However this is not the case. From eq.(\ref{QzbbLog}), we learn that $\nabla_zR_{\bar{z}i\bar{z}j}$ contains odd numbers of the extrinsic curvatures. Thus, $\frac{\partial L}{\partial \nabla_zR_{\bar{z}i\bar{z}j}}$ also flips the sign under $n_{\mu}\to -n_{\mu}$. It turns out $\frac{\partial L}{\partial \nabla_zR_{\bar{z}i\bar{z}j}}  K_{\bar{z}ij}$ as a whole is orientation independent. While for the convariant formula eq.(\ref{Waldsix2}), it should be stressed that $ K_{\beta\rho\sigma}=n_{\beta} ^{\ a} K_{a \rho\sigma}$ includes only the spacetime indexes and thus is actually orientation independent. So the generalized Wald entropy is indeed orientation independent.

Let us go on to study the anomaly of entropy. Because the general case is quite complicated we consider some special conical metrics below. For simplicity, we keep only the highest order of $T, V_i, Q_{ij}$ eq.(\ref{TVQ}) which is also the case studied in \cite{Dong, Camps}.

Recall that the squashed conical metric is
\begin{eqnarray}\label{conesix}
ds^2=e^{2A}[dzd\bar{z}+e^{2A}T(\bar{z}dz-zd\bar{z})^2]+2i e^{2A}V_i(\bar{z}dz-zd\bar{z})dy^i
+( g_{ij}+Q_{ij})dy^idy^j.
\end{eqnarray}
For simplicity, we firstly consider the case with zero extrinsic curvature. Thus, we have
\begin{eqnarray}\label{TVQsix}
&&T=T_1+T_{a}x^a+T_{ab}x^ax^b+...,\nonumber\\
&&V_i=U_i+V_a x^a+V_{abi}x^ax^b+...,\nonumber\\
&&Q_{ij}=Q_{abij}x^ax^b+....
\end{eqnarray}
Note that there is a factor $e^{2A}$ before $T_{z\bar{z}},V_{z\bar{z}i}$ and $Q_{z\bar{z}ij}$.
Let us calculate $R, \nabla R$, and select all the possible terms relevant to HEE. We have
\begin{eqnarray}\label{RDR}
&&R_{z\bar{z}z\bar{z}}=e^{2A}\partial_z\partial_{\bar{z}}A+...,\nonumber\\
&&\nabla_zR_{z\bar{z}z\bar{z}}=e^{2A}\partial_z^2\partial_{\bar{z}}A+...,\nonumber\\
&&\nabla_zR_{zizj}=4Q_{zzij}\partial_zA+...,\nonumber\\
&&\nabla_zR_{z\bar{z}zj}=-3i e^{2A}V_{zj}\partial_zA+....
\end{eqnarray}
Note that to derive $\nabla_zR_{zizj}$ and $\nabla_zR_{z\bar{z}zj}$, we have identified $ z\partial_z^2A\cong -\partial_zA$ and $ z^2\partial_z^3A\cong  2\partial_zA$. In general, we have $ z^m\partial_z^{m+1}A\cong (-1)^m m!\partial_zA$. We can read out these indentities from eq.(\ref{dAdA}).
Using eqs.(\ref{dA},\ref{dAdA},\ref{RDR}), we can derive HEE for six-derivative gravity as
\begin{eqnarray}\label{HEEsix1}
S_{HEE}&=&2\pi\int d^dy \sqrt{g}\big{[}\frac{\delta L}{\delta R_{z\bar{z}z\bar{z}}}+64\big{(}\frac{\partial^2L}{\partial \nabla_zR_{zizl}\partial \nabla_{\bar{z}}R_{\bar{z}k\bar{z}l}}\big{)}_{\alpha_1}\frac{Q_{zzij}Q_{\bar{z}\bar{z}kl}}{\beta_{\alpha_1}}\nonumber\\
&+&96 i\big{(}\frac{\partial^2L}{\partial \nabla_zR_{zizl}\partial \nabla_{\bar{z}}R_{\bar{z}z\bar{z}k}}\big{)}_{\alpha_1}\frac{Q_{zzij}V_{\bar{z}k}}{\beta_{\alpha_1}}+c.c\nonumber\\
&+&144\big{(}\frac{\partial^2L}{\partial \nabla_zR_{z\bar{z}zl}\partial \nabla_{\bar{z}}R_{\bar{z}z\bar{z}k}}\big{)}_{\alpha_1}\frac{V_{zl}V_{\bar{z}k}}{\beta_{\alpha_1}}
\big{]}.
\end{eqnarray}
Here $V_{ai}=\frac{1}{6}\epsilon^{\mu\nu}n^{\rho}_{\ a} g^{\sigma}_{\ i} R_{\mu\nu\rho\sigma}$, $T=\frac{1}{64}\epsilon^{\mu\nu}\epsilon^{\rho\sigma}R_{\mu\nu\rho\sigma}$. Note that we only need the traceless part of $Q_{abij}$ in the above formula. We denote the traceless part of $Q_{abij}$  by $\hat{Q}_{abij}=H_{abij}-\frac{1}{2}n_{ab}H_{c\ ij}^{\ c}$ with $H_{abij}=K_{am(i}K_{|b|j)}^{\ \ \ m}-n^{\mu}_{\ a}n^{\nu}_{\ b}g^{\rho}_{\ (i}g^{\sigma}_{\ j)}R_{\mu\rho\mu\sigma}$.
Let us rewrite the above formula in convariant form. We have
\begin{eqnarray}\label{Anomalysixconvaiant}
S_{HEE}&=&2\pi \int d^{d}y\sqrt{g}\Big[-\frac{\delta L}{\delta R_{\mu \nu \lambda \sigma}}\epsilon^{\mu \nu} \epsilon^{\lambda \sigma}+
4\Big(\frac{\partial^2 L}{\partial \nabla_{\mu_1}R_{\mu_2 \mu_3 \mu_4 \mu_5} \partial \nabla_{\nu_1}R_{\nu_2\nu_3\nu_4\nu_5}}\Big)_{\alpha} \frac{Q_{\lambda_1 \lambda_2 \mu_3 \mu_5}Q_{\lambda_3 \lambda_4 \nu_3 \nu_5}}{\beta_{\alpha}}\nonumber \\
&&[(n_{\mu_1 \nu_1}-i\epsilon_{\mu_1 \nu_1})(n_{\mu_2 \nu_2}-i\epsilon_{\mu_2 \nu_2})(n_{\mu_4 \nu_4}-i\epsilon_{\mu_4 \nu_4})(n^{\lambda_1
\lambda_3}-i\epsilon^{\lambda_1 \lambda_3})(n^{\lambda_2 \lambda_4}-i\epsilon^{\lambda_2 \lambda_4})]-\nonumber \\
&&2\Big(\frac{\partial^2 L}{\partial \nabla_{\mu}R_{\mu_2 \nu_3 \mu_4 \mu_5} \partial \nabla_{\nu_1} R_{\nu_2 \nu_3 \nu_4 \nu_5}}\Big)_{\alpha}\frac{Q_{\lambda_1 \lambda_2 \mu_3 \mu_5}R_{\lambda_4 \lambda_5 \lambda_6 \nu_5}}{\beta_\alpha}[(n_{\mu_1 \nu_1}-i\epsilon_{\mu_1 \nu_1})\nonumber \\
&&(n_{\mu_2 \nu_2}-i\epsilon_{\mu_2 \nu_2})(n_{\mu_4 \nu_4}-i\epsilon_{\mu_4 \nu_4})(n^{\lambda_1 \lambda_4}-i\epsilon^{\lambda_1 \lambda_4})(n^{\lambda_2 \lambda_6}-i\epsilon^{\lambda_2 \lambda_6})(n_{\nu_3}^{\ \lambda_5}-i\epsilon_{\nu_3}^{\ \lambda_5})]+c.c\nonumber \\
&&+\Big(\frac{\partial^2 L}{\partial \nabla_{\mu_1} R_{\mu_2 \mu_3 \mu_4 \mu_5}\partial \nabla_{\nu_1} R_{\nu_2 \nu_3 \nu_4 \nu_5}}\Big)_\alpha \frac{R_{\lambda_1 \lambda_2 \lambda_3 \mu_5}R_{\lambda_4 \lambda_5 \lambda_6 \nu_5}}{\beta_\alpha}[(n_{\mu_1 \nu_1}-i\epsilon_{\mu_1 \nu_1})\nonumber \\
&&(n_{\mu_2 \nu_2}-i\epsilon_{\mu_2 \nu_2})(n_{\mu_4 \nu_4}-i\epsilon_{\mu_4 \nu_4})(n^{\lambda_1 \lambda_4}-i\epsilon^{\lambda_1 \lambda_4})(n^{\lambda_2 \lambda_5}-i\epsilon^{\lambda_2 \lambda_5})(n^{\lambda_3 \lambda_6}-i\epsilon^{\lambda_3 \lambda_6})]\Big]
\end{eqnarray}
with
\begin{eqnarray}
Q_{\mu \nu \lambda \sigma}=n^{a}_\mu n^{b}_\nu g^i_\lambda g^j_\sigma \hat{Q}_{abij}
\end{eqnarray}
Note that we use Dong's notation $\epsilon_{z\bar{\bar{z}}}=\frac{i}{2}$. So the above formula is real although including $i$. To keep the expression simple, we do not expand it as product of $\epsilon_{\mu\nu}$ and $n_{\mu\nu}$. The final expression is not expected to include $i$ explicitly. It is not hard to proof the terms of the product which include odd number of $\epsilon_{\mu\nu}$ are vanishing.
Let's take the last one in (\ref{Anomalysixconvaiant}) an example. We make the index swap, $\mu_i \leftrightarrow \nu_i$ (i=1,2,3,4) and $\lambda_j \leftrightarrow \lambda_{j+3}$(j=1,2,3). $\Big(\frac{\partial^2 L}{\partial \nabla_{\mu_1} R_{\mu_2 \mu_3 \mu_4 \mu_5}\partial \nabla_{\nu_1} R_{\nu_2 \nu_3 \nu_4 \nu_5}}\Big)_\alpha \frac{R_{\lambda_1 \lambda_2 \lambda_3 \mu_5}R_{\lambda_4 \lambda_5 \lambda_6 \nu_5}}{\beta_\alpha}$ will keep the same. But the terms of
the products of $\epsilon_{\mu\nu}$ and $n_{\mu\nu}$ which include odd number $\epsilon_{\mu \nu}$ will give an extra minus sign. So these terms must be vanishing. The second one in (\ref{Anomalysixconvaiant}) include its complex partner, which also makes the terms including odd number of $\epsilon_{\mu \nu}$ vanishing.

Now let us consider a more complicated case. We set $V_i=0$ but with general $T,\ Q_{ij}$. For simplicity, we only investigate a special action, $S=\int dx^D\sqrt{G} \nabla_{\mu}R_{\nu\alpha \beta \gamma }\nabla^{\mu}R^{\nu\alpha \beta \gamma }$. Applying the formulae in the appendix, we obtain the anomaly of entropy
\begin{eqnarray}\label{sixderivatieHEEQ}
 S_{\text{Anomaly}}&=&32\pi\int d^dy\sqrt{ g}  \Big[4Q_{\bar z \bar z ij}Q_{zz}^{\ \ ij}+8 K_{\bar z }^{\ ij}K_{zj}^{\ \ k}Q_{z\bar z ki}
 -2K_{\bar zp}^{\ \ p}K_{\bar z }^{\ ij}Q_{zzij}\nonumber \\&-&2K_{zp}^{\ \ p}K_{z }^{\ ij}Q_{\bar z\bar zij}+2K_{zij}K_{\bar z}^{\ ij}Q_{z\bar zp}^{\ \ \ p}+(K_{zij}K_{\bar z}^{\ ij})^2+K_{zp}^{\ \ p}K_{\bar z q}^{\ \ q}K_{zij}K_{\bar z}^{\ ij}\nonumber \\
 &-&4K_{z}^{\ ij}K_{\bar z jk}K_{\bar z }^{\ kl}K_{zli}+ \tri_s^{(y)}K_{zij}\tri^{(y)s}K_{z}^{\ ij}-40TK_{zij}K_{\bar z}^{\ ij}\nonumber \\
 &+&K_{zij}K_{z}^{\ ij}K_{\bar z kl}K_{\bar z }^{\ kl}+R_{zijk}R_{\bar z }^{\ ijk}-6Q_{zz\bar zij}K_{\bar z}^{\ ij}-6Q_{\bar z \bar z zij}K_{z}^{\ ij}\nonumber \\
 &+&4 Q_{0z\bar zij}Q_{0z\bar z}^{\ \ \ ij}+36T_0^2-28T_0 tr(K_z K_{\bar z})-2Q_{0z\bar zij}K_{\bar z}^{\ il}K_{zl}^{\ \ j}+Q_{0z\bar zi}^{\ \ i}K_{\bar z j}^{\ j}K_{zl}^{\ \ l}\Big]
 \end{eqnarray}
As a check of formula, we shall use the above formula to derive the universal terms of EE for 4d CFTs in sect. 4.

To summary, we have found a new type of Wald entropy, the generalized Wald entropy, for six-derivative gravity. This generalized Wald entropy appears on entangling surfaces without the rotational symmetry and reduces to Wald entropy for stationary black holes. Here `entangling surface' denotes the co-dimension 2 surface where we calculate HEE in the bulk. `without the rotational symmetry' means we do not have the $U(1)$ symmetry alone the Euclidean time. Instead, we only have a discrete $Z_n$ symmetry ($n=1$ for entanglement entropy and $n$ is a positive integer for Renyi entropy). It would be interesting to study the physical meaning of this generalized Wald entropy. We leave it to the future work. We also derive the anomaly of entropy for cone metrics with zero extrinsic curvature. As for non-zero extrinsic curvature, we  study a toy model of six-derivative gravity. In sect. 4, we shall also prove that our results give the correct logarithmic term of EE for 4d CFTs.

\subsection{HEE for 2n-derivative gravity}

We calculate HEE of 2n-derivative gravity in this subsection.  By 2n-derivative gravity, we mean the equations of motion are 2n-order differential equations. Its action can always be rewritten as $S(g, R, \nabla R,..., \nabla^{n-2} R)$. In general, the formula of HEE becomes more and more complicated when higher and higher derivatives are involved. For simplicity, we consider only one special case here.

We choose the cone metric (\ref{cone}) with
\begin{eqnarray}\label{TVQ2n}
&&T=z^{n-3}T_{\underbrace{z...z}_{n-3}}+\bar{z}^{n-3}T_{\underbrace{\bar{z}...\bar{z}}_{n-3}}+\sum_{m=n-2}^{\infty} e^{2A P_{a_1...a_m}}T_{a_1...a_m}x^{a_1}...x^{a_m},\nonumber\\
&&V_i=z^{n-2}V_{\underbrace{z...z}_{n-2}i}+\bar{z}^{n-2}V_{\underbrace{\bar{z}...\bar{z}}_{n-2}i}+\sum_{m=n-1}^{\infty} e^{2A P_{a_1...a_m}}V_{a_1...a_m i}x^{a_1}...x^{a_m},\nonumber\\
&&Q_{ij}=z^{n-1}Q_{\underbrace{z...z}_{n-1}ij}+\bar{z}^{n-1}Q_{\underbrace{\bar{z}...\bar{z}}_{n-1}ij}+\sum_{m=n}^{\infty} e^{2A P_{a_1...a_m}}Q_{a_1...a_m ij}x^{a_1}...x^{a_m}.
\end{eqnarray}
We call this kind of cone as `the highest-order cone'. That is because only the highest-order derivative of curvature $\nabla^{n-2}R$ contributes to the anomaly of entropy in this case. We have
\begin{eqnarray}\label{DnR}
&&\nabla_z^{n-2}R_{zizj}=(n-1) \Gamma[n] Q_{\underbrace{z...z}_{n-1}ij}\partial_zA+...,\nonumber\\
&&\nabla_z^{n-2}R_{z\bar{z}zj}=- i \frac{n-2}{n-1}\Gamma[n+1]  e^{2A}V_{\underbrace{z...z}_{n-2}j}\partial_zA+...,\nonumber\\
&&\nabla_z^{n-2}R_{z\bar{z}z\bar{z}}= \frac{n-3}{n-2}\Gamma[n+1]  e^{4A}T_{\underbrace{z...z}_{n-3}}\partial_zA+....\nonumber\\
\end{eqnarray}
In the derivation of the above fromulas, we have identified $ z^m\partial_z^{m+1}A$ with $ (-1)^m m!\partial_zA$, which can be read out from eq.(\ref{dAdA}).

Using eqs.(\ref{dA},\ref{dAdA},\ref{DnR}), we can derive HEE for 2n-derivative gravity as
\begin{eqnarray}\label{HEEsix}
S_{EE}&=&2\pi\int d^dy \sqrt{g}\big{[}\frac{\delta L}{\delta R_{z\bar{z}z\bar{z}}}+4(n-1)^2\Gamma[n]^2\big{(}\frac{\partial^2L}{\partial \nabla_z^{n-2}R_{zizj}\partial \nabla_{\bar{z}}^{n-2}R_{\bar{z}k\bar{z}l}}\big{)}_{\alpha_1}Q_{\underbrace{z...z}_{n-1}ij}Q_{\underbrace{\bar{z}...\bar{z}}_{n-1}kl}/\beta_{\alpha_1}\nonumber\\
&+&i 8(n-2)\Gamma[n]\Gamma[n+1] \big{(}\frac{\partial^2L}{\partial \nabla_z^{n-2}R_{zizj}\partial \nabla_{\bar{z}}^{n-2}R_{\bar{z}z\bar{z}k}}\big{)}_{\alpha_1}Q_{\underbrace{z...z}_{n-1}ij}V_{\underbrace{\bar{z}...\bar{z}}_{n-2}k}/\beta_{\alpha_1}+c.c\nonumber\\
&+& 4\frac{(n-1)(n-3)}{n-2}\Gamma[n]\Gamma[n+1] \big{(}\frac{\partial^2L}{\partial \nabla_z^{n-2}R_{zizj}\partial \nabla_{\bar{z}}^{n-2}R_{\bar{z}z\bar{z}z}}\big{)}_{\alpha_1}Q_{\underbrace{z...z}_{n-1}ij}T_{\underbrace{\bar{z}...\bar{z}}_{n-3}}/\beta_{\alpha_1}+c.c\nonumber\\
&+&16\frac{(n-2)^2}{(n-1)^2}\Gamma[n+1]^2\big{(}\frac{\partial^2L}{\partial \nabla_z^{n-2}R_{z\bar{z}zl}\partial \nabla_{\bar{z}}^{n-2}R_{\bar{z}z\bar{z}k}}\big{)}_{\alpha_1}V_{\underbrace{z...z}_{n-2}l}V_{\underbrace{\bar{z}...\bar{z}}_{n-2}k}/\beta_{\alpha_1}\nonumber\\
&+& -i8\frac{(n-3)}{n-1}\Gamma[n+1]^2 \big{(}\frac{\partial^2L}{\partial \nabla_z^{n-2}R_{z\bar{z}zi}\partial \nabla_{\bar{z}}^{n-2}R_{\bar{z}z\bar{z}z}}\big{)}_{\alpha_1}V_{\underbrace{z...z}_{n-2}i}T_{\underbrace{\bar{z}...\bar{z}}_{n-3}}/\beta_{\alpha_1}+c.c\nonumber\\
&+&4\frac{(n-3)^2}{(n-2)^2}\Gamma[n+1]^2\big{(}\frac{\partial^2L}{\partial \nabla_z^{n-2}R_{z\bar{z}z\bar{z}}\partial \nabla_{\bar{z}}^{n-2}R_{\bar{z}z\bar{z}z}}\big{)}_{\alpha_1}T_{\underbrace{z...z}_{n-3}}T_{\underbrace{\bar{z}...\bar{z}}_{n-3}}/\beta_{\alpha_1}
\big{]}.
\end{eqnarray}

As for the general case, the formula of HEE is quite complicated. Like the holographic Weyl anomaly, it seems very difficult ( if not impossible ) to derive an exact expression. Actually, there is no need to work it out exactly. Instead, for any given action and cone metric, we can directly use eqs.(\ref{dA},\ref{dAdA}) to calculate HEE.

\section{Checks of our formulas}

In this section, we prove that our formula of HEE yields the correct logarithmic term of EE for 4d CFTs. This is a nontrivial check of our results. For simplicity, we focus on an example of 6-derivative gravity in five-dimensional space-time as follows:
\begin{eqnarray}\label{actionderivative}
S&=&\frac{1}{16\pi}\int
d^{5}x\sqrt{-\hat{G}}(\hat{R}+\frac{12}{l^2}+\lambda_1\nabla_{\mu}\hat{R}\nabla^{\mu}\hat{R}+\lambda_2\nabla_{\alpha}\hat{R}_{\mu\nu}\nabla^{\alpha}\hat{R}^{\mu\nu}+\lambda_3\nabla_{\alpha}\hat{R}_{\mu\nu\rho\sigma}\nabla^{\alpha}\hat{R}^{\mu\nu\rho\sigma}).\nonumber\\
\end{eqnarray}
According to \cite{Solodukhin} , the expected logarithmic term of EE for the dual CFTs is
\begin{eqnarray}\label{4dEE1}
S_{EE}=\log(l/\delta)\frac{1}{2\pi}\int_{\Sigma}
d^2x\sqrt{h}[aR_{\Sigma}-c(C^{abcd}h_{ac}h_{bd}-k^{\iota ab}k_{\iota
ab}+\frac{1}{2}k^{\iota a}_ak_{\iota b}^b) ],
\end{eqnarray}
where the central charges $a$ and $c$ is given by \cite{Miao}
\begin{eqnarray}\label{acderivative}
a=\frac{\pi}{8}, \ \ \
c=\frac{\pi}{8}+8\pi\lambda_3.
\end{eqnarray}
Thus, it is expected that HEE of $\nabla_{\mu}\hat{R}\nabla^{\mu}\hat{R}$ and $\nabla_{\alpha}\hat{R}_{\mu\nu}\nabla^{\alpha}\hat{R}^{\mu\nu}$ do not contribute to the logarithmic term, while HEE of  $\nabla_{\alpha}\hat{R}_{\mu\nu\rho\sigma}\nabla^{\alpha}\hat{R}^{\mu\nu\rho\sigma}$ yields a logarithmic term as
\begin{eqnarray}\label{4dEE}
-4\lambda_3\log(l/\delta)\int_{\Sigma}
d^2x\sqrt{h}[C^{abcd}h_{ac}h_{bd}-k^{\iota ab}k_{\iota
ab}+\frac{1}{2}k^{\iota a}_ak_{\iota b}^b ].
\end{eqnarray}
As we shall prove below, this is indeed the case.

Let us firstly compute the generalized Wald entropy.  Applying the formula (\ref{Waldsix2}), we get
\begin{eqnarray}\label{GWaldLog}
S_{\text{G-Wald}}=\frac{1}{4}\int d\rho d^2y \sqrt{h}\big{[}&&-1+2\lambda_1 \Box \hat{R}+\lambda_2 n^{ \mu\nu}\Box \hat{R}_{\mu\nu}+\lambda_3 \epsilon^{\mu\nu}\epsilon^{\rho\sigma}\Box \hat{R}_{\mu\nu\rho\sigma}\nonumber\\
&&+2\lambda_2 \nabla^{\alpha}\hat{R}^{\mu\nu}  K_{\beta} (n^{\beta}_{\ \mu}n_{\alpha\nu}-\epsilon^{\beta}_{\ \mu}\epsilon_{\alpha\nu})\nonumber\\
&&+2\lambda_3 \nabla^{\alpha}\hat{R}^{\mu\rho\nu\sigma}  K_{\beta\rho\sigma} (n^{\beta}_{\ \mu}n_{\alpha\nu}-\epsilon^{\beta}_{\ \mu}\epsilon_{\alpha\nu})\ \big{]}.
 \end{eqnarray}
Note that we work in the Euclidean signature. So HEE is different from the Lorentzian one by a minus sign. The first term of the above equation is just the Bekenstein-Hawking entropy. According to \cite{Solodukhin, Hung}, it gives a logarithmic term as
\begin{eqnarray}\label{EinsteinLog}
\log(l/\delta)\frac{1}{16}\int_{\Sigma}
d^2x\sqrt{h}[R_{\Sigma}-(C^{abcd}h_{ac}h_{bd}-k^{\iota ab}k_{\iota
ab}+\frac{1}{2}k^{\iota a}_ak_{\iota b}^b) ].
\end{eqnarray}
Thus we only need to consider the other terms of eq.(\ref{GWaldLog}) below.

For asymptotically AdS space-time, we can expand the bulk metric in the Fefferman-Graham gauge
\begin{eqnarray}\label{FGgauge}
ds^2=\hat{G}_{\mu\nu}dx^{\mu}dx^{\nu}=\frac{1}{4\rho^2}d\rho^2+\frac{1}{\rho}g_{ij}dx^idx^j,
\end{eqnarray}
where $g_{ij}=\overset{\scriptscriptstyle{(0)}}{g}_{ij}+\rho \overset{\scriptscriptstyle{(1)}}{g}_{ij}+...+\rho^{\frac{d}{2}}(
\overset{\scriptscriptstyle{(\frac{d}{2})}}{g}_{ij}+\overset{\scriptscriptstyle{(\frac{d}{2})}}{h}_{ij}\log \rho)+...$. Interestingly,
\begin{eqnarray}\label{g1}
\overset{\scriptscriptstyle{(1)}}{g}_{ij}=-\frac{1}{d-2}(\overset{\scriptscriptstyle{(0)}}{R}_{ij}-\frac{\overset{\scriptscriptstyle{(0)}}{R}}{2(d-1)}\overset{\scriptscriptstyle{(0)}}{g}_{ij}),
\end{eqnarray} can be determined completely by PBH transformation \cite{Theisen,Theisen1} and thus is independent of equations of motion. However, the higher order terms $\overset{\scriptscriptstyle{(2)}}{g}_{ij},\overset{\scriptscriptstyle{(3)}}{g}_{ij}...$  are indeed constrained by equations of motion. Fortunately, for the logarithmic terms of HEE in 5-dimensional space-time, we only need to expand the metric to the subleading order $\overset{\scriptscriptstyle{(1)}}{g}_{ij}$. Let us define a useful quantity $\tilde{R}$ as
\begin{eqnarray}\label{hatR}
&&\tilde{R}_{\mu\nu\rho\sigma}=\hat{R}_{\mu\nu\rho\sigma}+(\hat{G}_{\mu\rho}\hat{G}_{\nu\sigma}-\hat{G}_{\mu\sigma}\hat{G}_{\nu\rho}),\nonumber\\
&&\tilde{R}_{\mu\nu}=\hat{R}_{\mu\nu}+d \hat{G}_{\mu\nu},\nonumber\\
&&\tilde{R}=\hat{R}+d(d+1).
\end{eqnarray}
According to \cite{Miao}, we have
\begin{eqnarray}\label{order2}
&&\tilde{R}\sim o(\rho^2),\ \ \tilde{R}_{ij}\sim o(\rho),\
\ \ \tilde{R}_{i\rho}\sim o(\rho), \ \ \tilde{R}_{\rho\rho}\sim o(1)\nonumber\\
&&\tilde{R}_{i\rho j \rho}\sim o(\frac{1}{\rho}),\ \ \tilde{R}_{\rho ijk}\sim o(\frac{1}{\rho})\nonumber\\
&&\tilde{R}_{ijkl}=\frac{C_{ijkl}}{\rho}.
\end{eqnarray}
Note that eq.(\ref{g1}) is used in the derivation of above equations.

Denote the transverse space of the squashed cone by $m$. The embedding of the 3-dimensional submanifold $m$ into 5-dimensional bulk is described by $X^{\mu}=X^{\mu}(\sigma^{\alpha})$, where $X^{\mu}=\{x^i, \rho\}$ are bulk coordinates and $\sigma^{\alpha}=\{y^a,\tau\}$ are coordinates on $m$. We choose a gauge
\begin{eqnarray}\label{gaugeLog1}
\tau=\rho,\ \ h_{a\tau}=0,
\end{eqnarray}
where $h_{\alpha\beta}$ is the induced metric on $m$.  Let us expand the embedding functions as
\begin{eqnarray}\label{embeddingfunctions}
X^i(\tau, y^i)=\overset{\scriptscriptstyle{(0)}}{X^i}(y^a)+\overset{\scriptscriptstyle{(1)}}{X^i}(y^a)\tau+...
\end{eqnarray}
Diffeomorphism preserving the FG gauge (\ref{FGgauge}) and above gauge (\ref{gaugeLog1}) uniquely fixes a transformation rule of the embedding functions $X^{\mu}(y^a,\tau)$ \cite{Theisen2}. From this transformation rule, we can identity $\overset{\scriptscriptstyle{(1)}}{X^i}(y^a)$ with $\frac{1}{4}k^i(y^a)$
\begin{eqnarray}\label{rules1}
\overset{\scriptscriptstyle{(1)}}{X^i}(y^a)=\frac{1}{4}k^i(y^a),
\end{eqnarray}
where $k^i$ is the trace of the extrinsic curvature of the entangling surface $\Sigma$ in the boundary where CFTs live. From eq.(\ref{embeddingfunctions}), we can derive the induced metric on $m$ as
\begin{eqnarray}\label{inducedhab1}
 h_{\tau\tau}=\frac{1}{4\tau^2}\Big{(}1+\frac{1}{4}\,
 k^ik^j\overset{\scriptscriptstyle{(0)}}{g}_{ij}\,\tau+\cdots\Big{)},\ \ \ \
 h_{ab}=\frac{1}{ \tau}\Big{(}\overset{\scriptscriptstyle{(0)}}{h}_{ab}+\overset{\scriptscriptstyle{(1)}}{h}_{ab}\,\tau+...\Big{)},
\end{eqnarray}
with
\begin{eqnarray}\label{inducedhab2}
 \overset{\scriptscriptstyle{(0)}}{h}_{ab}=\partial_a \overset{\scriptscriptstyle{(0)}}{X^i}\partial_b \overset{\scriptscriptstyle{(0)}}{X^j}\ \overset{\scriptscriptstyle{(0)}}{g}_{ij},\ \ \ \overset{\scriptscriptstyle{(1)}}{h}_{ab}=\overset{\scriptscriptstyle{(1)}}{g}_{ab}-\frac{1}{2}k^ik^j_{ab} \overset{\scriptscriptstyle{(0)}}{g}_{ij}.
\end{eqnarray}
Thus, we have
\begin{eqnarray}\label{Deth}
\sqrt{h}=\sqrt{\overset{\scriptscriptstyle{(0)}}{h}}\frac{1}{2\rho^2}+....
\end{eqnarray}

Using eq.(\ref{embeddingfunctions}), we can also derive the extrinsic curvature $K$ of $m$ as
\begin{eqnarray}\label{hatK}
K^i_{ab}=(k^i_{ab}-\frac{k^i}{2}\overset{\scriptscriptstyle{(0)}}{h}_{ab})+...
\end{eqnarray}
Note that all the other components of $K^{\mu}_{\alpha\beta}$ are higher order terms which do not contribute to the logarithmic terms.

Now let us begin to derive the logarithmic term from the generalized Wald entropy eq.(\ref{GWaldLog}). Note that $\Box\sim o(1)$ and $(\epsilon^{\mu\nu}, n^{\mu\nu})$ take the same order as $G^{\mu\nu}$. Applying  eqs.(\ref{order2},\ref{Deth},\ref{hatK}), we find that, in addition to the Bekenstein-Hawking entropy, only
$\epsilon^{\mu\nu}\epsilon^{\rho\sigma}\Box R_{\mu\nu\rho\sigma}\sim o(\rho)$ in the generalized Wald entropy eq.(\ref{GWaldLog}) contribute to the logarithmic terms. After some calculations, we can derive
\begin{eqnarray}\label{GWaldLog1}
S_{\text{G-Wald}}&=&\frac{1}{4}\int d\rho d^2y \sqrt{h}\big{[}\lambda_3 \epsilon^{\mu\nu}\epsilon^{\rho\sigma}\Box \hat{R}_{\mu\nu\rho\sigma}+...\big{]}\nonumber\\
&=&\frac{1}{4}\int d\rho d^2y \frac{\sqrt{\overset{\scriptscriptstyle{(0)}}{h}}}{2\rho^2}\big{[}\lambda_3 \epsilon^{ij}\epsilon^{kl}(4\rho^2\nabla_{\rho}\nabla_{\rho}\hat{R}_{ijkl}+\rho \overset{\scriptscriptstyle{(0)}}{g}^{mn}\nabla_{m}\nabla_{n}\hat{R}_{ijkl})+...\big{]}\nonumber\\
&=&\frac{1}{4}\int d\rho d^2y \frac{\sqrt{\overset{\scriptscriptstyle{(0)}}{h}}}{2\rho}\big{[}\lambda_3 \overset{\scriptscriptstyle{(0)}}{\epsilon} ^{ij}\overset{\scriptscriptstyle{(0)}}{\epsilon} ^{kl}(-8 C_{ijkl})+...\big{]}\nonumber\\
&=&-4\int d\rho d^2y \frac{\sqrt{\overset{\scriptscriptstyle{(0)}}{h}}}{2\rho}\big{[}\lambda_3 ( h^{ik} h^{jl}C_{ijkl})+...\big{]}\nonumber\\
&=&-4\lambda_3 \log(l/\delta)\int d^2y \sqrt{\overset{\scriptscriptstyle{(0)}}{h}} ( h^{ac} h^{bd}C_{abcd})+...\nonumber\\
\end{eqnarray}
It agrees with the expected logarithmic term of EE for 4d CFTs with zero extrinsic curvature eq.(\ref{4dEE}).
In the above derivations, we have used the following useful formulae
\begin{eqnarray}\label{logformula}
\nabla_{\rho}\nabla_{\rho}\hat{R}_{ijkl}=\frac{C_{ijkl}}{\rho^3},\  \ g^{mn}\nabla_{m}\nabla_{n}\hat{R}_{ijkl}=-12\frac{C_{ijkl}}{\rho^2},\ \ \overset{\scriptscriptstyle{(0)}}{\epsilon} ^{ij}\overset{\scriptscriptstyle{(0)}}{\epsilon} ^{kl} C_{ijkl}=2h^{ac} h^{bd}C_{abcd}, \ \ \rho_0=\delta^2.
\end{eqnarray}

Now let us go on to compute the logarithmic term from the entropy eq.(\ref{sixderivatieHEEQ}). It should be mentioned that the splittings $T_0, Q_{0\ z\bar{z}ij}$ do not affect the discussions of this section. From eqs.(\ref{sixderivatieHEEQ},\ref{AnomalyL1},\ref{AnomalyL2},\ref{AnomalyL3}), $T_0, Q_0$ contribute terms in the form of $T_0^2,T_0 K^2, Q_0^2, Q_0 K^2$.  However, since $T_0\sim Q_0\sim K^2$, these terms are all of oder $O(K^4)$ which do not contribute to  logarithmic term of EE in 4d at all.  Recall that the squashed conical metric is
\begin{eqnarray}\label{coneLog}
ds^2=e^{2A}[dzd\bar{z}+e^{2A}T(\bar{z}dz-zd\bar{z})^2]+2i e^{2A}V_i(\bar{z}dz-zd\bar{z})dy^i
+( g_{ij}+Q_{ij})dy^idy^j
\end{eqnarray}
with
\begin{eqnarray}\label{TVQLog}
&&T=T+T_{a}x^a+...,\nonumber\\
&&V_i=U_i+V_a x^a+V_{abi}x^ax^b+...,\nonumber\\
&&Q_{ij}=2K_{aij}x^a+Q_{abij}x^ax^b+Q_{abcij}x^ax^bx^c....
\end{eqnarray}
Note that there is a factor $e^{2A}$ before $Q_{z\bar{z}},Q_{zz\bar{z}}, Q_{z\bar{z}\bar{z}}$ and $V_{z\bar{z}}$. It should be stressed that, for asymptotically AdS space-time the submanifold $m$ is very close to the boundary, thus we cannot choose $T, V_i, Q_{ij}$ freely. Instead, they should approach to the value for AdS. On the leading order, we have
\begin{eqnarray}\label{TVQLog1}
T=-\frac{1}{12}, \ U_i=0, \ V_{ai}=0, \ Q_{z\bar{z}ij}=\frac{1}{2}G_{ij}, \ Q_{zzij}=K_{zil}K_{zj}^{\ l}, \ Q_{zz\bar{z}ij}=\frac{4}{9}K_{zij}.
\end{eqnarray}
Let us derive the above formulas. For simplicity, we focus on pure AdS below. It is expected that it gives the leading value of $T, V, Q$ for asymptotically AdS.

According to \cite{Dong}, we have
\begin{eqnarray}\label{RDong}
\hat{R}_{abcd} &=& 12 T \varepsilon_{ab} \varepsilon_{cd}, \nonumber\\
\hat{R}_{abci} &=& 3 \varepsilon_{ab} V_{ci}, \nonumber\\
\hat{R}_{abij} &=& 2  \varepsilon_{ab} (\partial_i U_j - \partial_j U_i) + G^{kl} (K_{ajk} K_{bil} - K_{aik} K_{bjl}), \nonumber\\
\hat{R}_{aibj} &=& [\varepsilon_{ab} (\partial_i U_j - \partial_j U_i) + 4 G_{ab} U_i U_j ] + G^{kl} K_{ajk} K_{bil} - Q_{abij}, \nonumber\\
\hat{R}_{ikjl} &=& r_{ikjl} + G^{ab} (K_{ail} K_{bjk} - K_{aij} K_{bkl}).
\end{eqnarray}
Comparing the above formula with $\hat{R}_{\mu\nu\rho\sigma}=-G_{\mu\rho}G_{\nu\sigma}+G_{\mu\sigma}G_{\nu\rho}$, we get
\begin{eqnarray}\label{RDong1}
 &&T=-\frac{1}{12},  \nonumber\\
&&V_{ci}=0,\nonumber\\
 &&(\partial_i U_j - \partial_j U_i)=0, \ G^{kl} (K_{ajk} K_{bil} - K_{aik} K_{bjl})=0 ,\nonumber\\
&&G_{ab}G_{ij}+ 4 G_{ab} U_i U_j + G^{kl} K_{ajk} K_{bil} - Q_{abij} =0,\nonumber\\
&&r_{ikjl} +G_{ij}G_{kl}-G_{il}G_{kj}+ G^{ab} (K_{ail} K_{bjk} - K_{aij} K_{bkl})=0.
\end{eqnarray}
Let us make a brief discussion. Since $F_{ij}=\partial_i U_j - \partial_j U_i=0$, we can always set $U_i=0$ locally. Since $K$ is in higher order, from the last equation above, we find $G_{ij}$ is the metric of $AdS_3$ on leading order. To derive the leading order of $Q_{zz\bar{z}ij}$, one need to compute $\nabla_{\bar{z}} R_{zizi}$. To leading order, we have
\begin{eqnarray}\label{QzbbLog}
\nabla_{\bar{z}} \hat{R}_{zizj}=-4T_0 K_{zij}+2 K_{zl(i}Q_{z\bar{z}j)}^{\ l}-3Q_{zz\bar{z}ij}+o(K^3)=0.
\end{eqnarray}
Taking into account $T=-\frac{1}{12}, Q_{z\bar{z}ij}=\frac{1}{2}G_{ij}$, we get $Q_{zz\bar{z}ij}=\frac{4}{9}K_{zij}+o(K^3)$. Now we can calculate the logarithmic term from the anomaly of entropy.

Without loss of generality, to the leading order, we can choose the regularized conical metric as
\begin{eqnarray}\label{AdSconeLog}
ds^2&=&e^{2A}[dzd\bar{z}-\frac{1}{12}e^{2A}(\bar{z}dz-zd\bar{z})^2]+\frac{1+e^{2A}z\bar{z}}{4\rho^4}d\rho^2\nonumber\\
&+&\frac{\eta_{ab}(1+e^{2A}z\bar{z})+\sqrt{\rho} \big{(}(2z+\frac{4}{3}zz\bar{z})\bar{k}_{zab}+(2\bar{z}+\frac{4}{3}z\bar{z}\bar{z})\bar{k}_{\bar{z}ab}\big{)} }{\rho}dy^ady^b
\end{eqnarray}
where we have replaced $K$ by $k$ by using eq.(\ref{hatK}) and $\bar{k}_{zab}=(k_{zab}-\frac{k_z}{2}h_{ab})$ is the traceless part of $k_{zab}$. Substituting the above squashed cone metric into eqs.(\ref{AnomalyL1},\ref{AnomalyL2},\ref{AnomalyL3}), we get
\begin{eqnarray}\label{AnomalyLog}
S_{Anomaly}&=&16\lambda_3\int d\rho d^2y \frac{1}{2\rho}\sqrt{h}\big{[} \bar{k}_{z ab}\bar{k}_{\bar{z}}^{ab}+o(\rho)\big{]}\nonumber\\
&=& 4\lambda_3 \log(l/\delta)\int_{\Sigma}
d^2x\sqrt{h}(k^{\iota ab}k_{\iota
ab}-\frac{1}{2}k^{\iota a}_ak_{\iota b}^b)
\end{eqnarray}

Combining eqs.(\ref{EinsteinLog},\ref{GWaldLog1},\ref{AnomalyLog}), we finally obtain the logarithmic term of HEE as
\begin{eqnarray}\label{HEELog}
S_{EE}=\log(l/\delta)\int_{\Sigma}
d^2x\sqrt{h}[(\frac{1}{16})R_{\Sigma}-(\frac{1}{16}+4\lambda_3)(C^{abcd}h_{ac}h_{bd}-k^{\iota ab}k_{\iota
ab}+\frac{1}{2}k^{\iota a}_ak_{\iota b}^b) ],
\end{eqnarray}
which exactly agrees with the CFT results eq.(\ref{4dEE}). Now we finish the proof.

\section{Resolution of the HMS puzzle}

Hung, Myers and Smolkin find that the logarithmic term of EE derived from the trace anomaly of 6d CFTs agrees with the holographic result for entangling surfaces with rotational symmetry. However, mismatch appears when the entangling surfaces have no rotational symmetry even if the extrinsic curvature vanishes \cite{Hung}. We clarify this problem in this section. After considering the anomaly of entropy from the higher-derivative term $C^{ijkl}\nabla^2 C_{ijkl}$, we resolve this problem successfully.

Let us first review the approach of calculating the logarithmic term of EE from the trace anomaly for 6d CFTs \cite{Hung,Myers}.
In six dimensions, the trace anomaly takes the following form
 \be
\langle\,T^i{}_i\,\rangle=\sum_{n=1}^3 B_n\, I_n + 2 A \, E_6,
\label{trace6}
 \ee
where $E_6$ is the Euler density and $I_i$ are conformal invariants defined by
 \bea
I_1&=&C_{k i j l}C^{i m n j}C_{m~~\,n}^{~~k l} ~, \qquad
I_2=C_{i j}^{~~k l}C_{k l}^{~~m n}C^{~~~i j}_{m n}~,\nonumber \\
I_3&=&C_{i k l m}(\nabla^2 \, \de^{i}_{j}+4R^{i}{}_{j}-{6 \over 5}\,R
\,
\de^{i}_{j})C^{j k l m}\,.\label{trace6x}
 \eea
According to \cite{Hung,Solodukhin,Myers}, the universal
logarithmic term of EE can be identified with HEE of the trace anomaly. For entangling surfaces with the rotational symmetry, only Wald entropy contribute to HEE of the trace anomaly (\ref{trace6}). Thus, we have
 \beq
S_\mt{EE} = \log(\ell/\delta)\, \int d^{4}x\sqrt{h}\,\left[\,2\pi\,
\sum_{n=1}^3 B_n\,\frac{\partial I_n}{\partial R^{i j}{}_{k
l}}\,\tilde{\veps}^{i j}\,\tilde{\veps}_{k l} +2\,A\,
E_{4}\,\right]_{\Sigma}\,,
 \label{Waldformula8z}
 \eeq
where
 \bea
  \frac{\partial I_1}{\partial R^{i j}{}_{k l}}\,\tilde{\veps}^{i
j}\,\tilde{\veps}_{k l} &=& 3 \left(C^{j m n k}\, C_{m~~n}^{~~i
l}\,\tilde\veps_{i j}\,\tilde\veps_{k l} - \frac{1}{4}\,
C^{iklm}\,C^{j}_{~kl m }\,\tilde{g}^{\perp}_{i j} + \frac{1}{20}\,
C^{ijkl}\,C_{ijkl}\right)
 \,,\nonumber \\
\frac{\partial I_2}{\partial R^{i j}{}_{k l}}\,\tilde{\veps}^{i
j}\,\tilde{\veps}_{k l} &=&3\left(C^{k l m n}\, C_{m n}^{~~~ i j}\,
\tilde\veps_{i j}\,\tilde\veps_{k l} -C^{iklm}\,C^{j}_{~kl m
}\,\tilde{g}^{\perp}_{i j} + \frac{1}{5}\, C^{ijkl}\,C_{ijkl}\right)
 \,,\label{items9} \\
\frac{\partial I_3}{\partial R^{i j}{}_{k l}}\,\tilde{\veps}^{i
j}\,\tilde{\veps}_{k l} &=&  2\left(\Box\, C^{i j k l}+ 4\,R^{i}{}_{m}
C^{m j k l}-\frac{6}{5}\, R\, C^{i j k l}\right) \tilde\veps_{i
j}\,\tilde\veps_{k l}- 4\, C^{i j k l}\,R_{ik}\,\tilde{g}^{\perp}_{jl}
 \nonumber \\
&& \qquad+ 4\,C^{iklm}\,C^{j}_{~kl m }\,\tilde{g}^{\perp}_{i
j}-\frac{12}{5}\, C^{ijkl}\,C_{ijkl} \, . \nonumber
 \eea
The above result can be reliably applied for entangling surfaces with rotational symmetry. However, Myers et al find that it is inconsistent with the holographic result for entangling surfaces with zero extrinsic curvature but without a rotational symmetry. Assuming the conditions
\begin{eqnarray}\label{MyersCondition}
K_{aij}=0, R_{abci}=3\epsilon_{ab}V_{ci}=0,
\end{eqnarray}
they derive the holographic result for Einstein gravity as
\be
S_{\text{HEE}}=\pi\log(\ell/\delta)\int_{\Sigma} d^4y \sqrt{h}[2\overset{\scriptscriptstyle{(2)}}{g^{\hat{i}}}_{\hat{i}}-\overset{\scriptscriptstyle{(1)}}{g}_{\hat{i}\hat{j}}\overset{\scriptscriptstyle{(1)}}{g}^{\hat{i}\hat{j}}+\frac{1}{2}(\overset{\scriptscriptstyle{(1)}}{g^{\hat{i}}}_{\hat{i}})^2]\label{gogo}
\ee
The mismatch between holographic result eq.(\ref{gogo}) and CFT result eq.(\ref{Waldformula8z})  becomes
 \bea\label{discrepancy}
\Delta S = -4\pi B_3\log(l/\de)\, \int_{\Sigma}d^4x\sqrt{h}&(&C_{m n}{}^{r
s }C^{m n k l} \tilde{g}^\perp_{s l}\tilde{g}^\perp_{r k} -C_{m n
r}{}^s C^{m n r l}\tilde{g}^\perp_{s l}
   \\
&&+ 2 C_{m}{}^n{}_r{}^s C^{mkrl} \tilde{g}^\perp_{n
s}\tilde{g}^\perp_{k l} - 2C_{m}{}^n{}_r{}^s C^{mkrl}
\tilde{g}^\perp_{n l}\tilde{g}^\perp_{k s})\,,
 \nonumber
 \eea
Although eq.(\ref{discrepancy}) is derived in the case of Einstein gravity, Myers et al argue that it can be applied to the general case.

Now let us discuss the origin of the mismatch. First of all, as argued by HMS, the holographic results are the correct ones. Thus, something goes wrong with the CFT results. As we shall show below, some contributions are ignored in the CFT calculations. Following the assumption eq.(\ref{MyersCondition}), we focus on the conical metric (\ref{conesix}) with $K_{aij}=V_{ai}=0$. According our formula eq.(\ref{HEEsix}), in addition to the Wald entropy, a new term proportional to $Q^2$ also contribute to HEE
\begin{eqnarray}\label{Anomalysix}
S&=&2\pi\int d^dy \sqrt{g}\big{[}\frac{\delta L}{\delta R_{z\bar{z}z\bar{z}}}+64\big{(}\frac{\partial^2L}{\partial \nabla_zR_{zizl}\partial \nabla_{\bar{z}}R_{\bar{z}k\bar{z}l}}\big{)}_{\alpha_1}\frac{Q_{zzij}Q_{\bar{z}\bar{z}kl}}{\beta_{\alpha_1}} \big{]},
\end{eqnarray}
when the derivative of curvature is included in the action. Since only $I_3$ (\ref{trace6x}) contains such terms $C^{ijkl}\nabla^2 C_{ijkl}$, so the mismatch $\Delta S$ should be proportional to $B_3$. This explains the proposal of Myers et al that $\Delta S\sim B_3$. Now let us calculate the contribute from $C^{ijkl}\nabla^2 C_{ijkl}\cong-\nabla_{m} C_{ijkl}\nabla^{m}C^{ijkl}$ exactly. Applying the formula (\ref{Anomalysix}), we can derive the contribution ignored in eq.(\ref{Waldformula8z}) as
 \bea\label{discrepancy1}
\Delta S_1 = 128\pi B_3\log(l/ \de)\, \int_{\Sigma}d^4x\sqrt{h}(\bar{Q}_{zzij}\bar{Q}_{\bar{z}\bar{z}}^{\   \ \ ij}),
 \eea
where $\bar{Q}_{abij}=Q_{abij}-\frac{Q_{ab}}{4}g_{ij}$ is the traceless part of $Q_{abij}$.

Substituted the cone metric (\ref{conesix}) with $K_{aij}=V_{ai}=0$ into eq. (\ref{discrepancy}), we obtain
 \bea\label{discrepancy2}
\Delta S = 128 \pi B_3\log(l/ \de)\, \int_{\Sigma}d^4x\sqrt{h}(\bar{Q}_{zzij}\bar{Q}_{\bar{z}\bar{z}}^{\   \ \ ij}),
 \eea
which is exactly the same as eq.(\ref{discrepancy1}). Thus taking into account the contributions from the higher-derivative term $C^{ijkl}\nabla^2 C_{ijkl}$, the CFT results exactly match the holographic ones.

It should be mentioned that after this work is finished, there appears two related papers \cite{Astaneh1,Astaneh2}. By applying the FPS regularization \cite{Solodukhin1}, they find that total derivatives may contribute to non-zero entropy and they propose to use the entropy from total derivatives to explain the HMS mismatch \cite{Astaneh1,Astaneh2}. However, it is found that actually the proposal of \cite{Astaneh1,Astaneh2} does not resolve the HMS puzzle \cite{Miao3}. For convenience of the reader, we briefly review work of \cite{Miao3} in the next section.

In this paper, we use the LM regularization \cite{Maldacena1, Dong} instead of the FPS regularization \cite{Solodukhin1}. As a result, the entropy of covariant total derivatives is zero. \cite{Dong2}. For example, by using eqs.(\ref{dA}, \ref{dAdA}), one can prove that the entropy of $\Box R$ and $\Box C_{ijkl}C^{ijkl}$ vanishes for the general conical metrics eq.(\ref{cone}) with arbitrary splitting. For the detailed discussions on the entropy of total derivatives, please refer to \cite{Dong2}.

\section{Comparison with APS's proposal for HMS puzzle}

In this section, we briefly review the work of \cite{Miao3} to show that the proposal of \cite{Astaneh1,Astaneh2} can not solve the HMS puzzle. For simplicity, we focus on Einstein gravity.  We calculate the entropy for all of the terms in the holographic Weyl anomaly by using the LM regularization \cite{Maldacena1, Dong} and the APS regularization \cite{Astaneh1,Astaneh2}, respectively. It turns out that only the LM regularization\cite{Maldacena1, Dong} can yield consistent results with the holographic ones.

In the holographic approach, the universal terms of EE for 6d CFTs dual to Einstein gravity is given by \cite{Hung}
\begin{eqnarray}\label{HoloLog}
S_{\text{HEE}}=\pi\log(\ell/\delta)\int_{\Sigma} d^4y \sqrt{h}[2\overset{\scriptscriptstyle{(2)}}{g^{\hat{i}}}_{\hat{i}}-\overset{\scriptscriptstyle{(1)}}{g}_{\hat{i}\hat{j}}\overset{\scriptscriptstyle{(1)}}{g}^{\hat{i}\hat{j}}+\frac{1}{2}(\overset{\scriptscriptstyle{(1)}}{g^{\hat{i}}}_{\hat{i}})^2].
\end{eqnarray}
The above formula applies to the case with zero extrinsic curvatures. For the general case, please see \cite{Miao2}. For simplicity, we focus on the following conical metric with zero extrinsic curvatures
\begin{eqnarray}\label{ConeHMSpuzzle}
ds^2=dr^2+r^2d\tau^2+(\delta_{ij}+2\tilde{H}_{ij}r^{2} \cos t \sin t) dy^idy^j.
\end{eqnarray}
Then the holographic universal term of EE eq.(\ref{HoloLog}) becomes
\begin{eqnarray}\label{HoloLogCone}
S_{\text{HEE}}=-\frac{\pi}{40}\log(\ell/\delta)\int_{\Sigma} d^4y \sqrt{h}[ 8 tr \tilde{H}^2-(tr \tilde{H} )^2 ].
\end{eqnarray}

The holographic Weyl anomaly for Einstein gravity is given by  \cite{Henningson}
\begin{eqnarray}\label{WeylanomalyEinstein1}
<T^i_{\ i}>=\frac{1}{32}\big{(} -\frac{1}{2}RR^{ij}R_{ij}+\frac{3}{50}R^3+R^{ij}R^{kl}R_{ikjl}-\frac{1}{5}R^{ij}\nabla_i\nabla_j R+\frac{1}{2}R^{ij}\Box R_{ij}-\frac{1}{20}R\Box R  \big{)}.
\end{eqnarray}
Note that the curvature in our notation is different from the one of \cite{Henningson} by a minus sign. In the field theoretical approach, the universal term of EE can be derived as the entropy of the Weyl anomaly \cite{Hung, Solodukhin}. Below we calculate the universal term of EE in the field theoretical approach by using the APS regularization \cite{Astaneh1,Astaneh2} and the LM regularization \cite{Maldacena1, Dong}, respectively.

Following \cite{Astaneh1,Astaneh2}, we regularize the conical metric eq.(\ref{ConeHMSpuzzle}) as
\begin{eqnarray}\label{ConeAPS}
ds^2=f_{n}(r)dr^2+r^2d\tau^2+(\delta_{ij}+2\tilde{H}_{ij}r^{2n} \cos t \sin t) dy^idy^j,
\end{eqnarray}
where $f_n=\frac{r^2+b^2 n^2}{r^2+b^2}$ and $\tau\sim \tau+2n \pi$. Using the above regularizaed conical metric, we can derive the total entropy of eq.(\ref{WeylanomalyEinstein1}) in the Lorentzian signature as
\begin{eqnarray}\label{APSLog}
S_{\text{APS}}=-\frac{\pi}{5}\int_{\Sigma} d^4y \sqrt{h}[ tr \tilde{H}^2 ],
\end{eqnarray}
which does not match the holographic result eq.(\ref{HoloLogCone}) at all. Please refer to \cite{Miao3} for the details of the calculations.

Applying the approach of \cite{Maldacena1, Dong}, we regularize the conical metric eq.(\ref{ConeHMSpuzzle}) as
\begin{eqnarray}\label{ConeDong}
ds^2=\frac{1}{(r^2+b^2)^{1-\frac{1}{n}}}(dr^2+r^2d\tau^2)+(\delta_{ij}+2\tilde{H}_{ij}r^{2} \cos t \sin t)dy^i dy^j,
\end{eqnarray}
with $\tau\sim \tau+2\pi$. From the above regularizaed conical metric, we can derive the total entropy of eq.(\ref{WeylanomalyEinstein1}) in the Lorentzian signature as
\begin{eqnarray}\label{MGLog}
S_{\text{MG}}=-\frac{\pi}{40}\int_{\Sigma} d^4y \sqrt{h}[ 8 tr \tilde{H}^2-(tr \tilde{H} )^2 ],
\end{eqnarray}
which exactly agrees with the holographic result eq.(\ref{HoloLogCone}). Please refer to \cite{Miao3} for the derivations of eq.(\ref{MGLog}).

Now it is clear that it is the proposal of this paper rather than the proposal of \cite{Astaneh1,Astaneh2} that can solve the HMS puzzle. This implies that the LM regularization \cite{Maldacena1, Dong} instead of the APS regularization \cite{Astaneh1,Astaneh2} is the correct apporach of regularization.

\section{Conclusions}

In this paper, we investigate HEE for the most general higher derivative gravity. In particular, we find a new class of generalized Wald entropy on entangling surfaces without the rotational symmetry. It appears in the general higher derivative gravity and reduces to Wald entropy on Killing horizon or on the entangling surface with the rotational symmetry. We also find all the possible would-be logarithmic terms which contribute to the anomaly of entropy. Combining the generalized Wald entropy and the anomaly of entropy together, we obtain a formal formula of HEE for the most general higher derivative gravity. We work out this formula exactly for $2n$-derivative gravity for some interesting conical metrics. We prove that our formula yields the correct universal term of entanglement entropy for 4d CFTs. This is a strong support of our results. As an important application of our formulae, we solve the HMS puzzle that the logarithmic term of entanglement entropy derived from Weyl anomaly of CFTs does not match the holographic result even if the extrinsic curvature vanishes. We find that such mismatch comes from the contributions of the derivative of the curvature. Taking into account such contributions carefully, we find that the CFT result match the holographic one exactly. Finally, we find that there is splitting problem in the derivations of HEE. The splitting problem can be fixed by using equations of motion for Einstein gravity. As for higher derivative gravity, how to fix the splitting problem is a non-trivial and open problem. We hope to address this problem in future. Fortunately, the splitting problem does not affect the main results of this paper. That is because the splitting problem only affects the entropy at least of order $O(K^4)$ ($K$ denotes the extrinsic curvatures) for all the examples studied in this paper.

\section*{Acknowledgements}

R. X. Miao is supported by Sino-Germann (CSC-DAAD) Postdoc Scholarship Program. W. Z. Guo is supported by  Postgraduate Scholarship Program of China Scholarship Council. We thank Miao Li for his encouragement and support. We are grateful to School of Astronomy and Space Science at Sun Yat-Sen University for hospitality where part of the work was done. We thank Stefan Theisen,  Ling-Yan Hung and Tadashi Takayanagi for valuable comments and discussions.

\appendix

\section{Useful formulas}

In the section, we list some formulas which would be useful for the calculations of HEE for six-derivative gravity. For simplicity, we ignore the splitting problems. In other words, we set $T_0=Q_0=0$.

\begin{eqnarray}
\tri_{\bar z} R_{zizj}&=&2K_{zij}\partial_{\bar z} \partial_z A-\Big[2\partial_z A K_{\bar z i}^{\ \ k}K_{zkj}\nonumber \\
&-&2\bar z \partial_z A  (K_{zi}^{\ \ k}(Q_{\bar z\bar z kj}-2K_{\bar z k}^{\ \ n}K_{\bar z n j})+(i\leftrightarrow j)\Big],
\end{eqnarray}
\begin{eqnarray}
\tri_s R_{zizj}&=&2\partial_zA (\tri^{y}_s K_{zij}+4iU_sK_{zij})+12i\bar z \partial_z A V_{\bar z s}K_{zij}\nonumber \\
 &+& \Big[2\bar z \partial_z A K_{zkj}\Big(2iU^kK_{\bar z si}+g^{kl}(\partial_i K_{\bar z ls}+\partial_s K_{\bar z lt}-\partial_l K_{\bar z si})-2 \gamma^n_{si}K_{\bar zn}^{\ \ k}\Big)+(i\leftrightarrow j)\Big],
 \end{eqnarray}
 \begin{eqnarray}
 \tri_z R_{z\bar z z i}&=& -3ie^{2A}\partial_z A V_{zi}\nonumber \\
 &+& 2ie^{2A}\bar z \partial_z A (K_{\bar zi}^{\ \ k}V_{zk}-K_{ zi}^{\ \ k}V_{\bar zk}),
 \end{eqnarray}
 \begin{eqnarray}
 \tri_{\bar z} R_{z \bar z z i}= -2iU_i e^{2A}\partial_z \partial_{\bar z} A,
 \end{eqnarray}
 \begin{eqnarray}
 \tri_i R_{z \bar z z j}&=&-2K_{zij} \partial_z \partial_{\bar z} A-2K_{\bar z i}^{\ \ k}K_{z kj} \partial_z A\nonumber \\
 &-&2\bar z \partial_z A K_{zkj}(Q_{\bar z \bar z li}g^{lk}-2K_{\bar z}^{\ kn}K_{\bar z ni}),
 \end{eqnarray}
 \begin{eqnarray}
 \tri_z R_{zi \bar z j}&=&-2\partial_z A K_{\bar z i}^{\ \ n}K_{z nj}\nonumber \\
 &-&2\bar z \partial_z A( K_{zj}^{\ \ n}Q_{z\bar z ni}-2K_{\bar z }^{\ mn}K_{\bar z ni}K_{zmj}),
 \end{eqnarray}
 \begin{eqnarray}
 \tri_z R_{z\bar z ij}&=&2\partial_z A K_{\bar zj}^{\ \ n}K_{zni}\nonumber \\
 &-&\bar z \partial_z A(K_{zi}^{\ \ n}Q_{\bar z \bar z nj}+K_{z}^{\ nl}K_{\bar z nj}K_{zli})-(i\leftrightarrow j)
 \end{eqnarray}
 \begin{eqnarray}
 \tri_z R_{zijk}&=&2\partial_z A(\partial_j K_{zki}+4iU_j K_{zik}+K_{zlj}\gamma^l_{ik})\nonumber \\
 &+&2\bar z \partial_z A\Big[3iK_{ij}V_{\bar z k}+2iU^l K_{zlk}K_{\bar z ij}-(i\leftrightarrow j)
 -2K_{zlj}K_{\bar z\ m}^{\ l}\gamma^m_{ik}\nonumber \\
 &+& K_{\bar zj}^{\ \ m}(\partial_i K_{ z km}+\partial_k K_{ z im}-\partial_m K_{ z ik})\Big]-(j\leftrightarrow k),
 \end{eqnarray}
 \begin{eqnarray}
 \tri_l R_{zijk}&=& 4e^{-2A}\partial_{ z}A (K_{\bar z lj}K_{zik}-K_{\bar z lk}K_{zij})\nonumber \\
 &+&4e^{-2A}\bar z \partial_z A (Q_{\bar z \bar z lk} K_{zij}-Q_{\bar z z lj}K_{zik}).
 \end{eqnarray}
 \begin{eqnarray}
 \tri_z R_{ikjl}&=&4e^{-2A}\partial_z A (K_{zij}K_{\bar z kl}+K_{\bar z ij}K_{zkl})\nonumber \\
 &-&4e^{-2A}\bar z \partial_z A (K_{zij}Q_{\bar z \bar z kl}+K_{zkl}Q_{\bar z \bar z ij})-(j\leftrightarrow k),
 \end{eqnarray}
 \begin{eqnarray}
 \tri_z R_{zizj}&=&2K_{zij}\partial_z \partial_z A +\partial_z A(4Q_{zzij}-8K_{zi}^{\ \ l}K_{zlj})\nonumber \\
&-&e^{2A}\bar z \partial_z A\Big[-24TK_{zij}+i(\partial_j V_{zi}+\partial_i V_{zj})\nonumber \\
 &-&2iU^l (\partial_j K_{ z li}+\partial_i K_{ z lj}-\partial_l K_{ z ji})-2iV_{zk}\gamma^k_{ij}\nonumber \\
 &-&8U^k(U_j K_{zki}+U_iK_{zkj})+4e^{-2A}(k_{zj}^{\ \ k}Q_{z\bar z ik}+K_{zi}^{\ \ l}Q_{z\bar z lj})\nonumber \\&-&2Q_{zz \bar zij}-16K_{zik }K_z^{\ lk}K_{zlj}\Big].
 \end{eqnarray}

Applying the above formulas and eqs.(\ref{dA},\ref{dAdA}), let us compute the anomaly of entropy for some toy models. We recover the contributions from $T_0$ and $Q_{0\ z\bar{z}ij}$ in these examples.

For $L_1=\tri_\mu R\tri^{\mu}R$, we obtain
\begin{eqnarray}\label{AnomalyL1}
S_{\text{Anomaly}}=32\pi\int dy^d\sqrt{g}\Big( 3tr(K_z K_{\bar z})-K_z K_{\bar z}+6 T_0-2Q_{0\ z \bar{z}}\Big)^2,
\end{eqnarray}
where $K_a \equiv g^{ij}K_{aij}$, $tr (K_a K_b)=g^{ij}K_{ail}K_{bj}^{\ \ l}$, $Q_{ab}=g^{ij}Q_{abij}$, etc.

For $L_2=\tri_\mu R_{\rho \nu}\tri^{\mu}R^{\rho \nu}$, we get
\begin{eqnarray}\label{AnomalyL2}
S_{\text{Anomaly}}=&&-4\pi\int dy^d\sqrt{g} \Big[40TK_z K_{\bar z}-6K_z^2 tr K_{\bar z}^2+2K_zK_{\bar z} tr(K_{z}K_{\bar z})-2Q_{z\bar z}K_z K_{\bar z}\nonumber \\
&&-K_{z}^2K_{\bar z}^2-8trK_{z}^2 tr K_{\bar z}^2 -8tr (K_zQ_{z\bar z}) K_{\bar z}+18 tr(K_{\bar z}^2 K_z)K_z\nonumber \\
&&-8K_{z}tr(Q_{\bar z\bar z}K_z)+8tr (K_z^2)Q_{\bar z \bar z}+4K_z^2Q_{\bar z\bar z}-4Q_{zz}Q_{\bar z\bar z}\nonumber \\
&&+4Q_{zz\bar z}K_{\bar z}-8(tr K_{z}K_{\bar z})^2-8tr(K_{z}K_z K_{\bar z}K_{\bar z})-8tr(K_z K_{\bar z}K_zK_{\bar z})\nonumber \\
&&-\tri_i^{(y)} K_z \tri^{(y)i} K_{\bar z}-2R_{zkij}R_{zk'i'j'}g^{kj}g^{k'j'}g^{ii'}+(z\leftrightarrow \bar z)\Big]\nonumber \\
&&+8\pi \int dy^d\sqrt{g} \Big[72 T_0^2-24T_0 Q_{0\ z\bar{z}}+ 4tr (Q_{0\ z\bar{z}}^2)+ 4Q_{0\ z\bar{z}}^2 \nonumber \\
&&-28 T_0 K_z K_{\bar{z}}+24 T_0 tr (K_zK_{\bar{z}})+11 Q_{0\ z\bar{z}}K_zK_{\bar{z}}-20Q_{0\ z\bar{z}}tr(K_zK_{\bar{z}})\Big].
\end{eqnarray}

For $L_3= \nabla_{\mu}R_{\nu\alpha \beta \gamma }\nabla^{\mu}R^{\nu\alpha \beta \gamma }$, we have
\begin{eqnarray}\label{AnomalyL3}
 S_{\text{Anomaly}}&=&32\pi\int d^dy\sqrt{ g}  \Big[4Q_{\bar z \bar z ij}Q_{zz}^{\ \ ij}+8 K_{\bar z }^{\ ij}K_{zj}^{\ \ k}Q_{z\bar z ki}
 -2K_{\bar zp}^{\ \ p}K_{\bar z }^{\ ij}Q_{zzij}\nonumber \\&-&2K_{zp}^{\ \ p}K_{z }^{\ ij}Q_{\bar z\bar zij}+2K_{zij}K_{\bar z}^{\ ij}Q_{z\bar zp}^{\ \ \ p}+(K_{zij}K_{\bar z}^{\ ij})^2+K_{zp}^{\ \ p}K_{\bar z q}^{\ \ q}K_{zij}K_{\bar z}^{\ ij}\nonumber \\
 &-&4K_{z}^{\ ij}K_{\bar z jk}K_{\bar z }^{\ kl}K_{zli}+ \tri_s^{(y)}K_{zij}\tri^{(y)s}K_{z}^{\ ij}-40TK_{zij}K_{\bar z}^{\ ij}\nonumber \\
 &+&K_{zij}K_{z}^{\ ij}K_{\bar z kl}K_{\bar z }^{\ kl}+R_{zijk}R_{\bar z }^{\ ijk}-6Q_{zz\bar zij}K_{\bar z}^{\ ij}-6Q_{\bar z \bar z zij}K_{z}^{\ ij}\nonumber \\
 &+&4 Q_{0z\bar zij}Q_{0z\bar z}^{\ \ \ ij}+36T_0^2-28T_0 tr(K_z K_{\bar z})-2Q_{0z\bar zij}K_{\bar z}^{\ il}K_{zl}^{\ \ j}+Q_{0z\bar zi}^{\ \ i}K_{\bar z j}^{\ j}K_{zl}^{\ \ l}\Big]
 \end{eqnarray}

Remarkably, the entropy from the splittings takes the form $T_0^2, T_0 K^2, Q_0^2, Q_0 K^2$. They are all in order $K^4$ due to the fact $T_0\sim Q_0\sim O(K^2)$. As a result, the splitting probem does not affect our results in  Sect. 4 and Sect. 5.

\end{document}